\documentclass[preprintnumbers,amsmath,amssymb]{revtex4}
\usepackage{nicefrac}
\usepackage{amssymb}
\usepackage{amsmath}
\usepackage{mathrsfs}
\usepackage{graphics}
\usepackage{epsfig}
\usepackage{graphicx}
\usepackage{subfigure}
\usepackage{bm}
\usepackage{nicefrac}

\def\beq{\begin{equation}}
\def\eeq{\end{equation}}
\def\baq{\begin{eqnarray}}
\def\eaq{\end{eqnarray}}
\def\be{\begin{equation}}
\def\ee{\end{equation}}
\def\bea{\begin{eqnarray}}
\def\eea{\end{eqnarray}}

\def\a{\alpha}
\def\b{\beta}

\def\d{\delta}

\def\r{\rho}

\def\mK{{\mathcal K}}

\def\d{{\rm d}}

\def\tr{{\rm tr}}

\newcommand{\ta}{{a}}
\newcommand{\tX}{{X}}

\newcommand{\aq}{\alpha^{2}}
\newcommand{\bq}{\beta^{2}}
\newcommand{\ha}[1]{h^{#1}}

\newcommand{\bright}{\begin{flushright}}
\newcommand{\eright}{\end{flushright}}
\newcommand{\bminip}{\begin{minipage}}
\newcommand{\eminip}{\end{minipage}}
\newcommand{\bcent}{\begin{center}}
\newcommand{\ecent}{\end{center}}

\begin{document}

%\today
\unitlength = 1mm
%\hfill{CERN-PH-TH/2012-???\qquad NORDITA-2012-???}
\title{Exact Solutions in Massive Gravity}
\author{Gianmassimo Tasinato$^{(1)}$, Kazuya Koyama$^{(1)}$, Gustavo Niz$^{(1,2)}$}

\affiliation{
~ \\
 $^{(1)}$ Institute of Cosmology $\&$ Gravitation, University of Portsmouth,
\\\hskip0.2cm %Dennis Sciama Building,
 Portsmouth, PO1 3FX, United Kingdom\\
%$(2)$ CERN, PH-TH Division, CH-1211, Gen\`eve 23, Switzerland\\
%$(3)$ NORDITA, SE-106 91, Stockholm, Sweden
$^{(2)}$ Departamento de F\'{\i}sica, Universidad de Guanajuato,\\
 DCI, Campus Le\' on, C.P. 37150, Le\' on, Guanajuato, M\' exico.
}%

\begin{abstract}
\noindent 
Massive gravity is a good theoretical laboratory to study modifications of General Relativity. The theory offers a concrete set-up to study models of dark energy, since it admits cosmological self-accelerating solutions in the vacuum, in which the size of the acceleration depends on the graviton mass. Moreover, non-linear gravitational self-interactions, in the proximity of a matter source, manage to mimic the predictions of linearised General Relativity, hence agreeing with solar-system precision measurements. In this article, we review our work in the subject, classifying, on one hand, static solutions, and on the other hand, self-accelerating backgrounds. For
what respects  static solutions we exhibit black hole configurations, together with other solutions that recover General Relativity near a source via the Vainshtein mechanism. For the self-accelerating solutions we describe a wide class of cosmological backgrounds, including an analysis of  their stability.
    \end{abstract}
\maketitle

\smallskip

%%%%%%%%%%%%%%%%%%%%%%% sect 1 %%%%%%%%%%%%%%%%%%%%%%%%%%
\section{Introduction}

Can the graviton have a mass? A
 graviton mass breaks the diffeomorphism invariance of General Relativity (GR), but has the advantage to
 potentially provide a theory of dark energy  that 
%  it
%and
%This question is important because a massive graviton
 %breaks the diffeomorphism i 
  %could
    explains,   in a technically natural way,  the
present day acceleration of our Universe. At 
large scales,  gravity is modified with respect to GR, and the theory admits cosmological accelerating solutions  in which the size of acceleration depends on the graviton mass.
% and controls    the behavior of long wavelength gravitons   leading to 
 % cosmological accelerating solutions in which the size of acceleration depends on the graviton mass.
   This way to explain cosmological acceleration is technically natural in the 't Hooft sense, because in the limit of graviton mass
   going to zero one recovers the full diffeomorphism invariance of GR: hence, corrections to the size of dark
   energy must be proportional
   to the (tiny) graviton mass itself.
  
  % of our universe.

Not aware of all possible consequences %relevance
of  massive gravity for cosmology,
  Fierz and Pauli (FP), back in 1939, started
   %the stage for
 %in the  far past
the theoretical
 study
 of massive
 gravity from a field theory perspective \cite{Fierz:1939ix}.
 % have been  made to answer it.
%
%Fierz and Pauli (FP) in 1939 \cite{Fierz:1939ix}
%attempts to theoretically  address these
%
 %The first attempts
%
 %question of the graviton
% mass
 %date back to the work by Fierz and Pauli (FP) in 1939 \cite{Fierz:1939ix}.
They considered a mass term for linear gravitational perturbations, which is uniquely determined by requiring the absence of ghost degrees of freedom. The mass term breaks the gauge (diffeomorphism) invariance of GR, leading to a classical graviton with five degrees of freedom, instead of the two found in GR. There have been intensive studies into what happens beyond the linearized theory of FP. In 1972, Boulware and Deser (BD) found a scalar ghost mode at the nonlinear level, the so called sixth degree of freedom in the FP theory \cite{Boulware:1973my}.  This issue has been re-examined using an effective field theory approach,  where gauge invariance is restored by introducing St\"uckelberg fields \cite{ArkaniHamed:2002sp}. In this language, the St\"uckelberg fields
physically play the role of the additional scalar and vector graviton polarizations. They 
acquire nonlinear interactions which contain more than two time derivatives, signaling the existence of a ghost \cite{ArkaniHamed:2002sp}. In order to construct a consistent theory, nonlinear terms should be added
to the FP model, which are tuned to remove the ghost order by order in perturbation theory.
Interestingly, this approach sheds light on another famous problem with FP massive gravity; due to contributions of the scalar degree of freedom, solutions in the FP model do not continuously connect to solutions in GR, even in the limit of zero graviton mass. This is known as the van Dam, Veltman, and Zakharov (vDVZ) discontinuity \cite{vanDam:1970vg, Zakharov:1970cc}.  Observations such as light bending in the solar system would exclude the FP theory, no matter how small the graviton mass is. In 1972, Vainshtein \cite{Vainshtein:1972sx} proposed a mechanism to avoid this conclusion; in the small mass limit, the scalar degree of freedom becomes strongly coupled and the linearized FP theory is no longer reliable. In this regime, higher order interactions, which are introduced to remove the ghost degree of freedom, should shield the scalar interaction and recover GR on sufficiently small scales.

Until recently, it was thought to be impossible to construct a ghost-free theory for massive gravity that is compatible with current observations \cite{Creminelli:2005qk, Deffayet:2005ys}.
% In this essay, we explore the consequences of a promising new development that may achieve this.
 Using an effective field theory approach,
  one can show that in  order to avoid the presence of a ghost,
 interactions   have to be chosen in such a way that the equations of motion for the scalar  and vector
component of the St\"uckelberg field  contains no more than two time derivatives. Recently, it was shown that there is a finite number of derivative interactions for scalar lagrangians
that give rise to second order differential equations. These are dubbed Galileon terms because of a symmetry under a constant shift of the scalar field derivative \cite{Nicolis:2008in}. Therefore, one expects that any consistent nonlinear completion of FP contains these Galileon terms, at least in an appropriate range of scales in which the scalar dynamics can be somehow isolated from the remaining degrees of freedom;
this is  the so-called decoupling limit \cite{ArkaniHamed:2002sp}. This turns out to be a powerful criterium for building higher order interactions with the desired properties. Indeed, following this route, de Rham,  Gabadadze and Tolley constructed a family of ghost-free extensions to the FP theory, which reduce to the Galileon terms in the decoupling limit. We refer to the resulting theory as
$\Lambda_3$ massive gravity \cite{drgt}.

\smallskip
In this article, we review our work to build and analyze exact solutions in $\Lambda_3$  massive gravity. As we have briefly explained,  %in massive gravity
non-linear effects play an essential role to characterize phenomenological consequences of this theory. 
% Non-linear contributions to 
%the field equations controlling the fields involved become important in regimes that are relevant. 
Then, the analysis of
   exact solutions of the equations of motion, obtained
 by imposing
 appropriate symmetries (spherical symmetry for static space-times, or homogeneity and isotropy for cosmological set-ups),
 make manifest, in idealized but representative situations, how the non-linear dynamics of the graviton
 degrees of freedom respond to the presence of a source, or, at very large scales,  to the  graviton mass
 itself. After all, looking back to the past,  we know that the knowledge of exact solutions of non-linear field equations have been of crucial importance to understand GR. The Schwarzschild solution lead to the discovery of the concept of black hole, and play
 an essential role for analyzing the dynamics of objects around massive sources in GR; and modern cosmology would be unthinkable without the use of Friedmann-Robertson-Walker solutions.  Exact solutions in massive gravity might lead to the discovery and understanding of new features and concepts in a theory of gravitation that can lead to important developments for our comprehension of gravitational interactions.

\smallskip

This review is organized as follows: in Section \ref{theory} we explicitly construct the $\Lambda_3$ massive gravity theory, while in Section \ref{sphsols} the most general Ansatz for spherically symmetric solution is introduced. This leads to two branches of static solutions: one exhibiting the Vainshtein mechanism, and the other representing a generalisation of Schwarzschild-(A)dS black holes. In Section \ref{cosmosec} we explore cosmological self-accelerating solutions  and their stability under perturbations. Finally, in Section \ref{futuresec} we conclude, also outlining possible directions for future research.
% {\bf GT to complete}

\section{Ghost-free massive gravity}\label{theory}

We begin with the covariant Fierz-Pauli mass term in four-dimensions, given by
\be\label{lagFP}
\mathcal{L}_{FP}=m^2\sqrt{-g}\;{\cal U}^{(2)},\qquad\qquad {\cal U}^{(2)}=\left(H_{\mu\nu}H^{\mu\nu}-H^2\right),
\ee
where $m$ is a parameter with units of mass and the tensor $H_{\mu \nu}$ is a covariantisation of the metric perturbations, namely
\be\label{H}
g_{\mu \nu} = \eta_{\mu \nu} +h_{\mu\nu}\,\equiv\,H_{\mu \nu}+\Sigma_{\mu\nu},\qquad\qquad
\mathrm{with}\qquad
\Sigma_{\mu\nu}\equiv\partial_\mu \phi^\alpha \partial_\nu \phi^\beta \eta_{\alpha \beta}.
\ee
The St\"uckelberg fields $\phi^\alpha$
are introduced to restore  reparametrisation invariance, hence
transforming as scalar from the point of view
of the physical metric \cite{ArkaniHamed:2002sp}. The internal metric $\eta_{\alpha \beta}$ corresponds to a  non-dynamical reference
 metric,  usually assumed to be Minkowski space-time.
%   Therefore, around flat space,  we can rewrite $H_{\mu \nu}$ as
% \bea\label{defhmn}
% H_{\mu \nu}&=&h_{\mu \nu}+\eta_{\beta \nu}\partial_\mu \pi^\beta+\eta_{\alpha \mu}\partial_\nu
% \pi^\alpha-
% \eta_{\alpha \beta} \partial_\mu \pi^\alpha \partial_\nu \pi^\beta, \nonumber\\
% &\equiv& h_{\mu \nu}-{\mathcal Q}_{\mu\nu}.
% \eea
% From now on, indices are raised/lowered with the dynamical metric $g_{\mu\nu}$,
%  unless otherwise
% stated.
% %For example,
% %$H^\mu_{\,\,\nu}\,=\,g^{\mu \rho} H_{\rho \nu}$.
%  Moreover, the Lagrangian (\ref{lagFP}) is invariant under coordinate transformations $x^{\mu} \to x^{\mu} + \xi^{\mu}$, provided
% $\pi^{\mu}$ transforms as
% \be\label{pitransf}
% \pi^{\mu} \to \pi^{\mu} + \xi^{\mu}.
% \ee
  The  dynamics of the Stuckelberg fields $\phi^\alpha$ are at the origin of the two features discussed in the introduction: the BD ghost excitation and the vDVZ discontinuity.
  With respect to the first issue, as noticed
  by Fierz and Pauli, one can remove the ghost excitation, to linear order in perturbations, by choosing the
  quadratic structure $H_{\mu\nu}H^{\mu\nu}-H^2$.
   When expressed in
    the St\"uckelberg field language, terms in the action
    are arranged in a such way to constraint one of the four Stuckelberg fields to be non-dynamical. However, when going beyond linear order, this constraint disappears, signaling  the emergence
     of an additional ghost mode \cite{ArkaniHamed:2002sp}. Remarkably, Ref.~\cite{deRham:2010ik} has shown how to
%     de Rham and Gabadadze  were able to
      construct a potential, tuned
     at each order in powers of $H_{\mu\nu}$,
      to hold the constraint and remove one of the St\"uckelberg fields. Even though the potential is expressed in terms of  an infinite series of terms for $H_{\mu\nu}$, it can be resummed into the following finite form \cite{drgt, Koyama:solutions}
     \be
{\cal U}= -m^2\left[{\cal U}_2+\alpha_3\, {\cal U}_3+\alpha_4\, {\cal U}_4\right],
\label{potentialU}
\ee
where $\alpha_n$ are free dimensionless parameters, ${\cal U}_n=n!\det_{n}(\mK)$ and the tensor ${\cal K}_{\mu}^{\ \nu}$ is defined as
\bea
{\mathcal K}_{\mu}^{\ \nu} &\equiv &\delta_{\mu}^{\ \nu}-\left(\sqrt{g^{-1} \Sigma}\right)_{\mu}^{\ \nu}.
\eea
(The square root is formally understood as $\sqrt{{\cal K}}_{\mu}^{\ \alpha}\sqrt{\cal K}_{\alpha}^{\ \nu}={\cal K}_{\mu}^{\ \nu}$.)  The relationship of these potentials with a determinant resides on the following property, which holds for squared real matrices and a complex number $z$ 
\be\label{detA}
\det\left( {\mathbb I}+ z \mK\right)=1+\sum_{n=1}^\infty z^n \det_n(\mK),
\ee
where each determinant can be written in terms of traces as
\bea
\det_{1}(\mK)&=&\tr\mK,\nonumber \\
\det_{2}(\mK)&=&(\tr\mK)^2-\tr (\mK^2),\nonumber \\
\det_{3}(\mK)&=&(\tr \mK)^3 - 3 (\tr \mK)(\tr \mK^2) + 2 \tr \mK^3,\nonumber \\
\det_{4}(\mK)&=& (\tr \mK)^4 - 6 (\tr \mK)^2 (\tr \mK^2)
+ 8 (\tr \mK)(\tr \mK^3) + 3 (\tr \mK^2)^2 - 6 \tr \mK^4\,. \nonumber
\eea
All terms $\det_{n} (\mK) $ with $n>4$ vanish in four dimensions. If one chooses a sum of determinants of the form $ \sum_{n=1}^4\det(\mathbb{I}+z_n\mK)-4$,
one can generate each $\det_n(\mK)$ term with a separate coefficient $\alpha_n$, provided a solution to $\sum_{i=1}^{4}z_n^i=\alpha_n$ exists, which is guaranteed by the Newton identities. Therefore, the massive gravity theory can be written in full as
\be\label{genlag}
 {\cal L} = \frac{M_{Pl}^2}{2}\,\sqrt{-g}\left( R -2\Lambda - m^2{\cal U}\right),
\ee
where ${\cal U}$ is given by (\ref{potentialU}) and we have introduced an
additional bare cosmological constant $\Lambda$. Notice that in order to obtain the Fierz Pauli term (\ref{lagFP}) as the first order correction to GR, we have ignored the tadpole term $\det_1(\mK)$.

We can write the equations of motion in a more familiar way by using the potential term (\ref{potentialU}) as the source of peculiar energy momentum tensor. In this way, the Einstein equations read
\be\label{einsteineqns}
G_{\mu \nu}=T^{{\cal U}}_{\mu \nu},
\ee
where the energy momentum tensor is defined as
\be
T^{{\cal U}}_{\mu \nu}\,=\frac{m^2}{\sqrt{-g}}\,\frac{ \delta \sqrt{-g}\ {\cal U}}{\delta g^{\mu \nu}}.
\ee
The theory defined by (\ref{genlag}) has Minkowski spacetime as trivial solution when $\Lambda=0$, hence one can rewrite the metric
$g_{\mu\nu}$ and the scalars $\phi^\mu$ as deviations from flat space, namely
\be\label{phi-pi}
g_{\mu\nu}=\eta_{\mu\nu}+h_{\mu\nu},\qquad \qquad \phi^\a=x^\a-\pi^\a,
\ee
where $x^\a$ are the usual cartesian coordinates spanning $\eta_{\alpha\beta}$. In what follows we will use $\phi^\alpha$ or $\pi^\alpha$, having in mind that (\ref{phi-pi}) relates them. Moreover, the {\it unitary}
gauge (where $\pi^\mu=0$) simplifies the potential (\ref{potentialU}) considerably, and we will start
with  this choice in what follows.
%{\bf GN I have restored the $\phi$ notation, since it was wrong to just call $\phi=\pi$ and I believe it is easier to include the $\phi$ again, instead of adding a $x^\mu$ terms everywhere, which becomes difficult to deal with in the decoupling limit}

There have been intensive studies in the issue of BD ghost in this theory \cite{ghost}. The general (but not universal
 \cite{Chamseddine:2013lid}) consensus is that there is indeed no BD ghost and the maximum number of propagation modes in this theory is five. However, this does not preclude a possibility that one of the five modes becomes a ghost around some backgrounds.

\section{Spherically Symmetric Solutions}\label{sphsols}
In this section, we review spherically symmetric solutions in the unitary gauge by following Refs.~\cite{Koyama:prl, Koyama:solutions}. 
The most general Ansatz with spherical symmetry, before fixing the gauge, is
\be\label{metricatz}
d s^2\,=\,-b(t,r)^2 \,d t^2+a(t,r)^2\, d r^2 +2 d(t,r)\, dt dr+c(t,r)^2 d \Omega^2,
\ee
where $d \Omega^2 = d \theta^2 + \sin^2 \theta d \phi^2$, and the Stuckelberg fields have the structure
% given by
\be\label{stuckatz}
\phi^0=f(t,r), \qquad \qquad \qquad
\phi^i=g(t,r)\frac{x^i}{r}.
\ee
%In the first
% part of this work
 We start  focussing  on the unitary gauge ($\pi^\mu=0$ or, equivalently, $f=t$ and $g=r$)
  and look for static solutions that do not depend, explicitly, on time.
%  {\bf GT in another part there will be Kazuya's section
 %on the review of W. Hu's approach }
  %It is important to stress
   %that solutions that are static in this unitary gauge
  The metric ansatz (\ref{metricatz}) reduces to
\be\label{genmetr}
d s^2\,=\,-b(r)^2 \,d t^2+a(r)^2\, d r^2 +2 d(r)\, dt dr+c(r)^2 d \Omega^2\,.
\ee
Furthermore, we choose to write the non-dynamical flat metric in (\ref{H}) as $ds^2 = -dt^2 + dr^2 + r^2 d \Omega^2$. It should be noticed that this is not a coordinate choice, but a way to simplify the expressions.
 Indeed,   we have chosen the unitary gauge, hence we are left with a theory which is  not 
   diffeomorphism invariant. Hence, in 
   this context  physics {\it does depend} on the choice of coordinates: we  will further explore this fact in section \ref{cosmosec}.  %, even for the fiducial, flat space
%Furthermore, as we will see in what follows, 
 Any change
of coordinate normally breaks the unitary gauge and switches on a non-trivial profile for the St\"uckelberg fields.
This also implies that the static solutions considered in the unitary gauge do not provide all the spherically
symmetric and static solutions in this theory. Other static solutions might exist with non-trivial St\"uckelberg fields turned on.

 Conscious of these limitations, let us start with this gauge choice: we will relax it in what follows. We plug the previous metric into the Einstein equations (\ref{einsteineqns}), and observe that  the Einstein tensor $G_{\mu\nu}$ satisfies the identity $d(r)\, G_{tt}+b(r)^2\,G_{tr}\,=\,0$, which
implies the algebraic constraint $0= d(r)\, T^{{\cal U}}_{tt}+b(r)^2\,T^{{\cal U}}_{tr}$. This last equation implies 
\be\label{conseq}
d(r)\left(c_0 r-c(r)\right)\,=\,0\,,
\ee
where $c_0$ is a function of $\alpha_3$ and $\alpha_4$ only (see Section \ref{branchII}). This constraint is solved in two possible ways, defining two branches of solutions:
% either the metric is diagonal
\be
d(r)\,=\,0 \hskip 0.9cm \text{or} \hskip 0.9cm c(r)\,=\,c_0\, r\,\,.%\, \hskip 0.9cm \text{with $c_0$ a constant }.
\ee
%Notice that this classification only holds in the unitary gauge, since one can always map the metric from one class to the other by a coordinate transformation, but to the price of exciting components of $\pi^\mu$.

In the next sections we will analyze each of these two branches separately. We will start from the diagonal one $d(r)=0$
 in Section \ref{branchI},  where the Vainshtein effect takes place and can be analyzed in a systematic way. Then we will proceed in Section \ref{branchII} to study  the
 class of solutions with a non-diagonal metric and  $c(r)\,=\,c_0\, r$, corresponding to non-asymptotically flat,  Schwarzschild-(Anti)-de Sitter
solutions that can be relevant to explain present-day cosmological acceleration.
% In this sector of the
%theory GR is recovered without appealing to  Vainshtein mechanism.
%However, in the other class of solutions, where the metric is diagonal in the unitary gauge,
%the situation is different. As we discuss in the next section, $\pi^\mu$ may or may not be strongly coupled; it could be strongly coupled within certain radial region, leading to a Vainshtein effect. This branch is the one that concerns us in the next Section.

\subsection{Branch I: Vainshtein mechanism at work}\label{branchI}

The problem of finding exact  vacuum solutions in this branch $d(r)=0$ is an open question, but we 
can make interesting progresses by considering perturbations (not necessarily small) from flat space. The following Ansatz is useful
\be
b(r)= 1+N(r) dt^2,\qquad a(r)= \left(1+F(r)\right)^{-1/2}\qquad c(r)=\left(1+H(r)\right)^{-1} \label{diagonal}.
\ee
Furthermore, it is convenient to introduce a new radial coordinate
$\rho = \frac{r}{1+H(r)}$, so that the linearised metric is expressed as
\be
ds^2 = - (1 + n) dt^2 + (1 - f) d\rho^2
+ \rho^2 d \Omega^2,
\ee
where $f(\rho) =  F\big(r(\rho)\big) - 2  h(\rho) - 2 \rho  h'(\rho)$, $n(\rho)=2N\big(r(\rho)\big)$, $h(\rho)=H\big(r(\rho)\big)$ and a prime denotes a derivative with respect to $\rho$.
As discussed above, one should be careful with this change of coordinates since, after fixing a
gauge, a change of frame in the metric
 breaks the unitary gauge and switches on
the St\"uckelberg fields $\pi^\mu$. It
turns out that this coordinate transformation excites the radial component of $\pi^\mu$, which
explicitly reads $\pi^\rho\,=\,\rho\, h$. Therefore, from now on one can think of $h$ as geometrically corresponding to  the
only non-zero component of the St\"uckelberg field $\pi^\mu$. At linear order, the equations for
the functions $n(\rho)$, $f(\rho)$ and $h(\rho)$ in the new variable radial $\rho$ are
\bea\label{Neq}
0&=& \left(m^2\rho^2+2\right)f+2 \rho \left(f'+m^2 \rho^2
h'+3 \, m^2 \rho h \right), \\
0&=&  \frac{1}{2}m^2 \rho^2 (n-4 h) -\rho \, n'-f, \label{Feq}
\\
0&=&  f +\frac{1}{2}\rho \, n'\label{const}.
\eea
In this linear expansion, the solutions for $n$ and $f$ are
\bea
n &=& - \frac{8 G M}{3 \rho} e^{- m \rho} \,,\label{linsoln}\\
f &=& -\frac{4 G M}{3 \rho} (1 + m \rho) e^{- m \rho}\,, \label{linsolf}
\eea
where we fix the integration constant so that $M$ is the mass of a point particle at the origin,
and $8 \pi G = M_{pl}^{-2}$. These solutions exhibit the vDVZ discontinuity, since the post-Newtonian
parameter $\gamma=f/n$ is $\gamma=\frac{1}{2}(1+m\rho)$, which in the massless limit reduces to $\gamma=1/2$,
in disagreement with GR predictions and solar system observations.

However, in order to understand what really happens in this limit, we must carefully analyse the behaviour
of the function $h(r)$ as $m\rightarrow 0$.  The study of the equations of motion for this metric
component makes manifest the non-linear effects responsible for the Vainshtein mechanism. To do this, we consider scales below the
Compton wavelength $m \rho \ll 1$, and at the same time ignore higher order terms in $G M$. Under
these approximations, the equations of motion can still be truncated to linear order in $f$
and $n$, but since $h$ is not necessarily small, we have to keep all non-linear terms in $h$. In
other words, we take the usual weak field limit for the metric fields, but keep all non-linearities
in the component $h$, since we expect regions where non-linear effects in $h$
become important.
% field is strongly coupled.
 As shown
in \cite{Koyama:solutions}, the field equations reduce to the following system of coupled equations for the fields $f$, $n$, $h$:
\bea
\label{feq}
f& =& - 2 \, \frac{G M}{\rho} - (m \rho)^2 \Big(
h - \alpha h^2 + \beta h^3 \Big) \,,\\
n' &=& 2 \, \frac{G M}{\rho^2} - m^2 \rho \, \Big( h - \b h^3 \Big)\label{neq} \,,\\
0&=&\frac{3}{2} \, \bq \, \ha{5}(\r) - \Big( \aq + 2 \b \Big) \, \ha{3}(\r) + 3 \, \Big( \a + \b A(\r) \Big) \, \ha{2}(\r)
- \frac{3}{2} \, h(\r) - A(\r) = 0 \label{heq}\,,
\eea
where
\bea
\a & \equiv& 1 + 3 \, \a_3\,\\
\b &\equiv& \a_3 + 4 \, \a_4\,,
\\
A(\r) & =& \big( \r_{v} / \r \big)^{3}\,,\\
\r_{v}&\equiv& \left( G M / m^{2} \right)^{\! 1/3}
\,.
\eea

%$\a \equiv 1 + 3 \, \a_3$ and $\b \equiv \a_3 + 4 \, \a_4 \,$, $A(\r) = \big( \r_{v} / \r \big)^{3}$ and $\r_{v}\equiv \left( G M / m^{2} \right)^{\! 1/3}$.
%As it
%was shown in \cite{Koyama:solutions}, these equations can also be obtained directly from the decoupling theory (see section \ref{declimitsec} for the definition of this limiting theory).
Equation (\ref{heq}) is a quintic algebraic equation in $h$, except for the special case where $\b = 0$, where it reduces to a cubic equation. 
% the two cases $\beta=0$ and $\beta\neq0$ must be treated separately since they can lead to different
%phenomenological consequences. 
Thus, after obtaining a solution for $h$ from equation (\ref{heq}), one can calculate the gravitational potentials $f$ and $n$ using (\ref{feq}) and (\ref{neq}).
In the particular case of $\beta=0$, it is possible to describe the solution in a simple way \cite{Koyama:prl}. For large radial $\rho$ values, one can linearise the equations in $h$, recovering the solution in Eqs.~(\ref{linsoln})-(\ref{linsolf}), to first order in $m \rho$. On the other hand, the Vainshtein mechanism applies, and below the so-called Vainshtein radius, $\rho_V = (G M m^{-2})^{1/3}$,
$h$ becomes larger than one and the non-linear terms in $h$ in eq. (\ref{heq}) become important, recovering GR close to a matter source. Actually, for $\rho \ll \rho_V$ the solution for $h$ is simply given by $|h| = \rho_V/(\a\rho) \gg 1$.
The latter solution for $h$ and Eq. (\ref{heq}) with $\beta=0$ implies
% {\bf GN I have included the $\a$ in the PRL solution below}
\bea
2 \rho n' &=& \frac{ 2 G M}{\rho} \left(1 + \frac{1}{2\a} \left(\frac{\rho}{\rho_V}
\right)^2 \right),\nonumber \\
f &=& - \frac{ 2 G M}{\rho} \left(1 -\frac{1}{2\a} \left(\frac{\rho}{\rho_V}
\right) \right).\label{solfbv}
\eea
Therefore, corrections to the GR solutions are indeed small for $\rho<\rho_V$, as shown in the left plot of Fig.~{\ref{plotprl}.  
Note that if we consider a finite size matter source, it was shown that there is no stable solution that interpolates from the Vainshtein region to the asymptotically flat solution,  and the Vainshtein region is naturally matched onto a solution that asymptotes to a non-flat cosmological background \cite{Berezhiani:2013dw}.% {\bf KK I rewrote this sentence}. 

%However, the $\beta=0$ solution presented here is not physical beyond the vacuum equations, because if one considers a matter source, then the outer solution to the source would match an asymptotically non-flat solution instead .

\begin{figure}[htp!]
\begin{center}
\includegraphics[width=8cm]{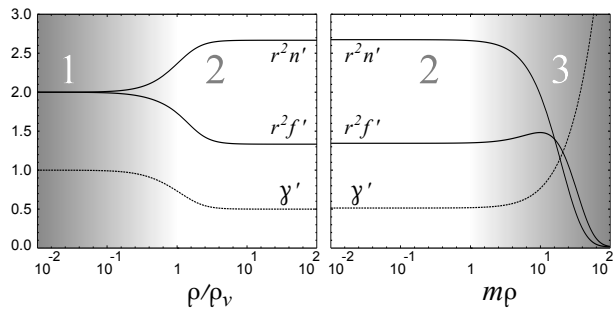}
\caption{Numerical solution for $\partial_r f=f'$, $\partial_r n=n'$, and the quotient $\gamma'\equiv f'/2n'$ around the Vainshtein radius $\rho_v$ (left) and the Compton Wavelength $\rho\sim 1/m$ (right). Region 1 (2) shows how GR solutions are (not) recovered inside (outside) the Vainshtein radius $\rho_V$. Region 3 shows the asymptotic decay of the linear solutions (Eqs.~(\ref{linsoln})-(\ref{linsolf})). Here,  $GM=1$ and $\a=1$.}
\label{plotprl}
\end{center}
\end{figure}

For the $\b\neq0$ case, analytic solutions of the previous algebraic equations can not be found in general. In the following we will follow the approach presented by Ref.~\cite{Koyama:Vainshtein}. It is possible to determine exactly how many local solutions exist in a neighbourhood of infinity at  $\r = +\infty$, which we refer as {\it asymptotic solutions}, and moreover how many local solutions exist in a neighbourhood of $\r = 0^{+}$, which we call {\it inner solutions}. Furthermore, we can find analytically
their leading behaviour as a function of $\rho$. Any global solution of (\ref{heq}) should necessarily interpolate between one of the asymptotic solutions and one of the inner solutions. Therefore, our aim is to understand, for each point in the  $(\a, \b)$ phase space, whether and how the above solutions match.

In a neighbourhood of $\r \to +\infty$ there are, depending on the value of $(\a, \b)$, three or five solutions
to eq.~(\ref{heq}). In particular, there is always a decaying solution, which we indicate with $\textbf{L}$. Its asymptotic behaviour is
$h(\r) \sim \left( \rho_{v}/\r\right)^{3} $. This solution corresponds to a spacetime which is
asymptotically flat. Additionally, there are two or four solutions to eq.~(\ref{heq}) which tend to a finite,
nonzero value as $\r \to +\infty$. We name these solutions with $\textbf{C}_{+}$, $\textbf{C}_{-}$,
$\textbf{P}_{1}$ and $\textbf{P}_{2}$ (details about this denomination are given in \cite{Koyama:Vainshtein}). Their asymptotic behaviour is  $h(\r) = C$, with $C$ a constant. These solutions correspond to spacetimes which are asymptotically non-flat. Interestingly, the leading term in the gravitational potentials scales as $\rho^2$ for large radii, the same scaling which we find in a de Sitter spacetime. It is worthwhile to point out that, since we are working on scales below the Compton wavelength of the gravitational field, {\it asymptotically non-flat} does not really mean the real behaviour at infinity. To understand the {\it true} asymptotic behaviour of this solution, one should solve the  complete, non-truncated equations.

In a neighbourhood of $\r \to 0^+$ there are either one or three solutions to eq.~(\ref{heq}).
For $\b > 0$ there are exactly three inner solutions, while for $\b < 0$ there is only one inner solution.
In particular,  there is always a diverging solution, which we denote by $\textbf{D}$. Its leading behaviour is
$h(\r) \sim  - \, \sqrt[3]{2/\b}\, (\r_v/\r) $. This solution exists for both $\b > 0$ and $\b < 0$, with opposite signs for each case. Using this solution in eqs.~(\ref{feq})-(\ref{neq}), one realises that the $h^3$ term cancels the $GM/\rho$ term, so the gravitational field is self-shielded and does not diverge as $\r \to 0^+$. This solution is in strong disagreement with gravitational observations. For $\b > 0$, there are two additional solutions to eq.~(\ref{heq}), which tend to a finite, non-zero value as $\r \to 0^+$. We indicate these solutions by $\textbf{F}_{+}$ and $\textbf{F}_{-} \,$, and their leading behaviour is $h(\r) = \pm (3 \, \b)^{-1/2}$. Notice that for $\b < 0$ there are no solutions to eq.~(\ref{heq}) which tend to a finite value as $\r \to 0^+$.
The expressions (\ref{feq})-(\ref{neq}) for the gravitational potentials imply that the metric associated to these solutions ($\textbf{F}_{+}$ and $\textbf{F}_{-}$) approximate the linearised Schwarzschild metric as $\r \to 0^+$.

From the behaviour of the inner solutions, one concludes that only in the $\b>0$ part of the phase space solutions may exhibit the Vainshtein mechanism \cite{Chkareuli:2011te}, but not necessarily for all values of $\alpha$ \cite{Koyama:Vainshtein}.  The phase space diagram which displays our results about solution matching is given in figure
\ref{phase space}. We discuss separately the $\b > 0$ and $\b < 0$ part of the phase space, and
refer to the figure for the numbering of the regions. The notation $\textbf{I}
\leftrightarrow \textbf{A}$ means that there is matching between the inner solution $\textbf{I}$
and the asymptotic solution $\textbf{A}$.
\begin{figure}[htp!]
\begin{center}
\includegraphics[width=7cm]{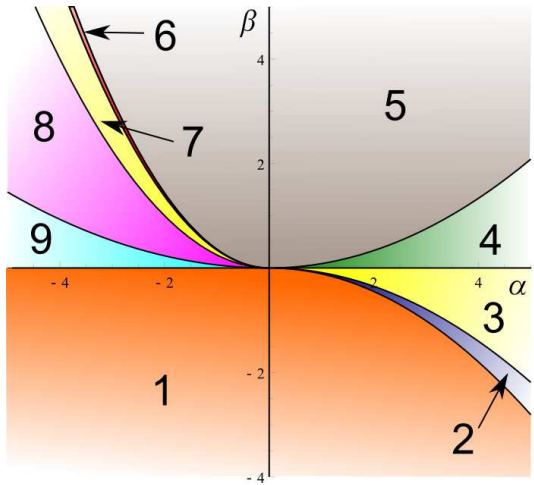}
\caption{Phase space diagram in $(\a,\b)$ for the solutions to the quintic equation
(\ref{heq}) in $h$, where the different regions show different matching of inner
solutions to asymptotic ones. The lines splitting the regions are half parabolas
($\b \propto \aq$, with $\a >0$ or $\a <0$) due to rescaling symmetry of eq.~(\ref{heq}).}
\label{phase space}
\end{center}
\end{figure}
\\
\\
\small
\textbf{$\b < 0$}
\normalsize
\\
In this part of the phase space, there is only one inner solution, $\textbf{D}$, so there can be at most one global solution to (\ref{heq}). There are three distinct regions which differ in the way the matching works (see \cite{Koyama:Vainshtein} for details):
\begin{itemize}
 \item[-] region 1: $\textbf{D} \leftrightarrow \textbf{C}_{+}$. The boundaries of this region are the line $\b = 0$ for $\a<0$ and the parabola $\b = c_{12} \, \a^2$ for $\a>0$, where $c_{12}$ is the negative root of the equation $-4 - 8 \, y + 88 \, y^2 - 1076 \, y^3 + 2883 \, y^4 = 0$. %(approximatively, $c_{12} \simeq -0.1124$). %On the boundary $\b = c_{12} \, \a^2$ the matching $\textbf{D} \leftrightarrow \textbf{C}_{+}$ still holds.

 \item[-] region 2: $\textbf{No matching}$. The boundaries of this region are the parabola $\b = c_{12} \, \a^2$ and the parabola $\b = c_{-} \, \a^2$, where $c_{-}$ is the only real root of the equation $8 + 48 \, y - 435 \, y^2 + 676 \, y^3 = 0$.% (approximatively, $c_{-} \simeq -0.0876$).

 \item[-] region 3: $\textbf{D} \leftrightarrow \textbf{P}_{2}$.
 \end{itemize}

\noindent \small
          \textbf{$\b > 0$}
          \normalsize
\\
In this part of the phase space, there are three inner solutions, $\textbf{D}$, $\textbf{F}_{+}$ and $\textbf{F}_{-}$, so there can be at most three global solutions to eq.~(\ref{heq}). There are six distinct regions with different matching properties:
\begin{itemize}
 \item[-] region 4: $\textbf{F}_{-} \leftrightarrow \textbf{L}$ , $\textbf{D} \leftrightarrow \textbf{C}_{-}$. The boundaries of this region are the parabola $\b = c_{45} \, \a^2$, where $c_{45} = 1/12$, and the line $\b = 0$. %On the boundary $\b = c_{45} \, \a^2$ there is the additional matching $\textbf{F}_{+} \leftrightarrow \textbf{C}_{+}$, and the corresponding solution is $h(\r) = const = + \sqrt{1/\,3\,\b}\,$.

 \item[-] region 5: $\textbf{F}_{+} \leftrightarrow \textbf{C}_{+}$ , $\textbf{F}_{-} \leftrightarrow \textbf{L}$, $\textbf{D} \leftrightarrow \textbf{C}_{-}$. The boundaries of this region are the parabola $\b = c_{45} \, \a^2$ for $\a > 0$ and the parabola $\b = c_{56} \, \a^2$ for $\a < 0$, where $c_{56} = (5 + \sqrt{13})/24$. %On the $\a < 0$ boundary $\b = c_{56} \, \a^2$ the matching works as in the rest of the region, but the solution $\textbf{F}_{-} \leftrightarrow \textbf{L}$ has an inflection point with vertical tangent.

 \item[-] region 6: $\textbf{D} \leftrightarrow \textbf{C}_{-}$ , $\textbf{F}_{+} \leftrightarrow \textbf{C}_{+}$. The boundaries of this region are the parabolas $\b = c_{56} \, \a^2$ and $\b = c_{67} \, \a^2$, where $c_{67}$ is the positive root of the equation $-4 - 8 \, y + 88 \, y^2 - 1076 \, y^3 + 2883 \, y^4 = 0$. %On the boundary $\b = c_{67} \, \a^2$ the matching works as in the rest of the region, but the solution $\textbf{D} \leftrightarrow \textbf{C}_{-}$ has an inflection point with vertical tangent.

 \item[-] region 7: $\textbf{F}_{+} \leftrightarrow \textbf{C}_{+}$. The boundaries of this region are the parabola $\b = c_{67} \, \a^2$ and the parabola $\b = c_{+} \, \a^2$, where $c_{+} = 1/4$. %Note that on the ($\a < 0$) part of the parabola $\b = 1/3 \, \aq$ there is the additional matching $\textbf{F}_{-} \leftrightarrow \textbf{C}_{-}$, so for these points there are two global solutions to eq.~(\ref{quintic}). On the boundary $\b = c_{+} \, \a^2$ there are the additional matchings $\textbf{F}_{-} \leftrightarrow \textbf{P}_{1}$ , $\textbf{D} \leftrightarrow \textbf{P}_{2}$, and the solutions corresponding to both of these additional matchings, seen as functions of $A$, display an infinite derivative in $A=0$.

 \item[-] region 8: $\textbf{F}_{+} \leftrightarrow \textbf{C}_{+}$ , $\textbf{F}_{-} \leftrightarrow \textbf{P}_{1}$ , $\textbf{D} \leftrightarrow \textbf{P}_{2}$. The boundaries of this region are the parabolas $\b = c_{+} \, \a^2$ and $\b = c_{89} \, \a^2$, where $c_{89} = (5 - \sqrt{13})/24$. %On the boundary $\b = c_{89} \, \a^2$ the matchings are the same as in the rest of the region, but the solution $h(\r)$ correspondent to the matching $\textbf{F}_{+} \leftrightarrow \textbf{C}_{+}$ has an inflection point with vertical tangent.

 \item[-] region 9: $\textbf{F}_{-} \leftrightarrow \textbf{P}_{1}$ , $\textbf{D} \leftrightarrow \textbf{P}_{2}$. %This region lies inside the $\a < 0$, $\b > 0$ part of the five roots at infinity region of the phase space (see fig.~\ref{five roots}), so there are again five asymptotic solutions. The matching is similar to that of region 8, apart from the fact that $\textbf{C}_{+}$ cannot be extended to $\r \to 0^+$ anymore; hence there are just two global solutions to eq.~(\ref{quintic}). The boundaries of this region are the parabola $\b = c_{89} \, \a^2$ and line $\b = 0$.
\end{itemize}

\noindent We note that the decaying solution $\textbf{L}$ never connects to the diverging configuration $\textbf{D}$,
so we can not have a spacetime which is asymptotically flat and exhibits the self-shielding of the
gravitational field at the origin. On the other hand, finite non-zero asymptotic solutions ($\textbf{C}_{\pm}$ or $\textbf{P}_{1,2}$) can connect to both finite and diverging inner solutions.
Therefore, one can have an asymptotically non-flat spacetime which presents self-shielding at the origin, or an asymptotically non-flat spacetime which tends to Schwarzschild spacetime for small radii. More precisely, for $\b < 0$ there are only solutions displaying the self-shielding of the gravitational field, apart from region 2 where there are no global solutions. Therefore the Vainshtein mechanism never works for $\b < 0$. In contrast, for $\b > 0$ all three kinds of global solutions are present. Solutions with asymptotic flatness and the Vainshtein mechanism are present in regions 4 and 5, while solutions which are asymptotically non-flat and exhibit the Vainshtein mechanism do exist in all ($\b > 0$) regions but region 4. Finally, solutions which display the self-shielding of the gravitational field are present in all ($\b > 0$) regions but region 7.

% \subsection{Numerical solutions}
%
% We present here the numerical solutions for the $h$ field and the gravitational potentials in some
% representative cases. We choose a specific realisation for each of the three physically distinct
% cases, namely asymptotic flatness with Vainshtein mechanism, asymptotically non-flat spacetime with Vainshtein mechanism, and asymptotically non-flat spacetime with self-shielded gravitational field at the origin. In addition, we consider the case in which there are no global solutions to eq.~(\ref{quintic}). This provides an illustration of what happens, in general, to local solutions of eq.~(\ref{quintic}) which cannot be extended to the whole radial domain, and give an insight on the phenomenology of equation (\ref{quintic}).
% \\
% \\
% \small
% \textbf{Asymptotic flatness with Vainshtein mechanism}
% \normalsize
% \\
% \\
For the sake of clearness, we show  one representative plot with the numerical matching solution between the inner and the asymptotic solutions.  We consider solutions which recover the Schwarzschild solution near the origin, and which are asymptotically flat ($\textbf{F}_{-} \leftrightarrow \textbf{L}$ in Figure \ref{Numerical1}-left), or non-flat ($\textbf{F}_{+} \leftrightarrow \textbf{C}_{+}$ in Figure \ref{Numerical1}-right). Finally, it is essential to decide whether these vacuum solutions are indeed consistent with matter sources, as it was done for the $\beta=0$ case in \cite{Berezhiani:2013dw}.
%In constrast, in Figure \ref{Numerical3}), we consider a self-shielding solution.

The Vainshtein mechanism has been studied intensively in the context of the Dvali-Gabadadze-Porrati braneworld model \cite{DGP} and Galileon models \cite{Vainshtein}. Especially, it was shown that the most general second order scalar tensor theory described by the Hordenski action \cite{Hordenski} leads to the same field equations as massive gravity \cite{Narikawa}.% {\bf KK I added this paragraph} 

\begin{figure}[htp!]
\begin{center}
\includegraphics[width=7cm]{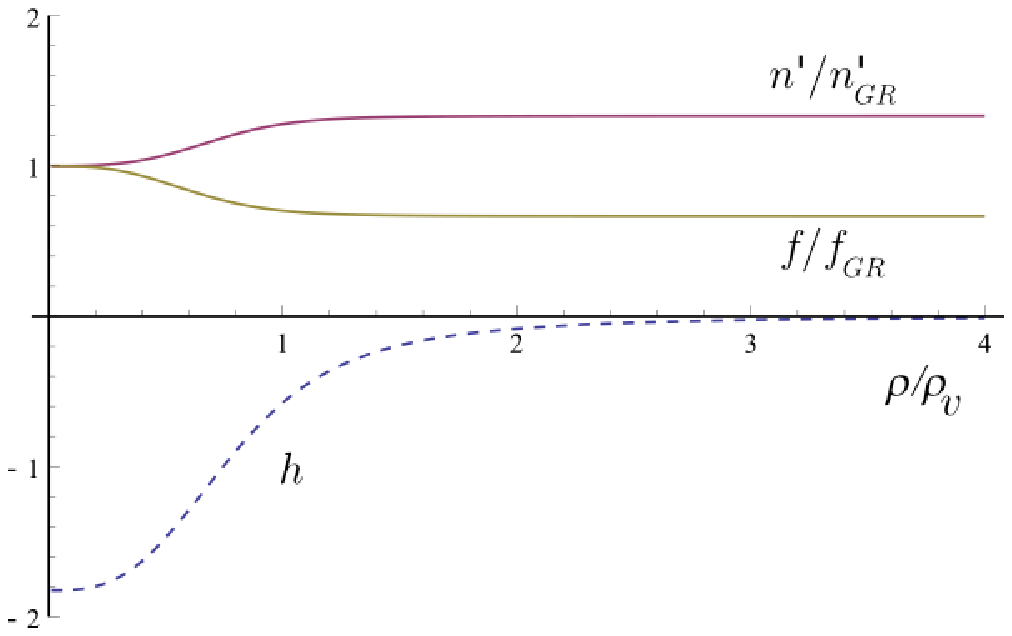}\hspace{1cm}
\includegraphics[width=7cm]{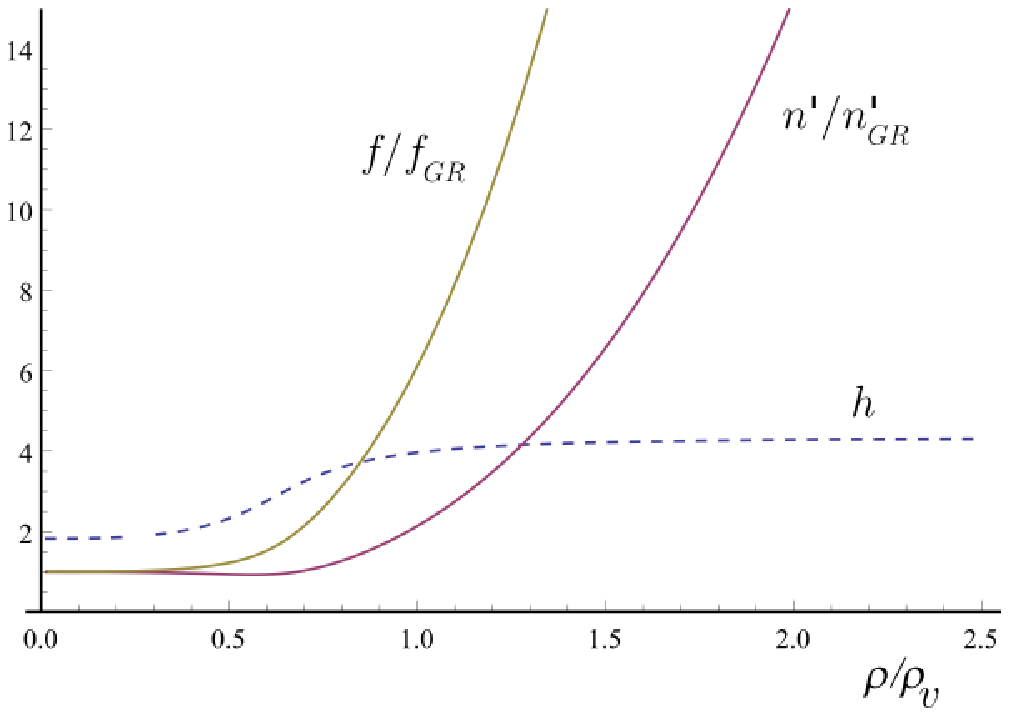}
\caption{Numerical solutions for $\textbf{F}_{-} \leftrightarrow \textbf{L}$ (left) and $\textbf{F}_{+} \leftrightarrow \textbf{C}_{+}$ (right), with $(\alpha, \beta) = (0 \, , 0.1)$. $f_{GR} = n_{GR} = - 2 G M / \r$. Solutions represent the GR gravitational potentials, and clearly present the vDVZ discontinuity with its resolution via the Vainshtein mechanism. Solutions asymptote a flat spacetime but with a different $\gamma=f/n$ than the Schwarzschild solution in GR (left) and a non-flat spacetime (right).}
\label{Numerical1}
\end{center}
\end{figure}

\subsection{Branch II: exact solutions}\label{branchII}

%{\bf GT add coupling to matter}

As we learned in the previous section, an essential property of this theory
of massive gravity
is the strong coupling phenomenon occurring
 in the proximity of a source.  On the other hand, the graviton mass
  induces non-linearities in the behavior of long wave length gravitons,
   responsible for the emergence of the second branch of solutions that we are going to study in this section.
   %
   %
   %cumulative effects
%
 %
 % This allows, for certain regions of parameter
%space, to recover linearised General Relativity at sufficiently small distances by
%means of the Vainshtein  mechanism.
%%In  unitary gauge, strong coupling manifests
%% with large contributions to the function $H(r) \,= \,1+h(r)$:
 % % the function $h(r)$
% %becomes much larger than one for distances smaller than the Vainshtein
 % %radius
 % %from the source.
   % This behaviour, accompanied
 %by  the fact that Birkhoff theorem
 % does not apply in this context,    suggests that exact solutions for this theory,
 % even imposing spherical symmetry, might be very different from the GR
  %ones.
%
 % In this section, we will exhibit new
 % spherically symmetric exact solutions in the vacuum for  massive gravity.
%   that generalize the ones of \cite{Salam:1976as}.
  In an appropriate gauge, these solutions are asymptotically de Sitter or Anti-de Sitter,
  depending on the choice of parameters.

We start with  the unitary gauge ($\pi^\mu=0$)
  and allow for arbitrary couplings
  $\alpha_3$ and $\a_4$, while from now on we set,  for simplicity, the bare cosmological constant to vanish (see \cite{Koyama:prl, Koyama:solutions} for a more complete discussion including a bare cosmological constant).
  We choose the static Ansatz  of eq. (\ref{genmetr})    
  for the metric, and we focus on the second branch of solutions for the constraint equation (\ref{conseq}): $c(r)\,=\,c_0\,r$. 
   Then, the exact solution of field equations is given by 
    \cite{Salam, Koyama:prl, Koyama:solutions},
  \bea
  c(r)&=& c_0\, r\,,\nonumber
  \\
  b(r)^2&=& b_0 +\frac{b_1}{r}+b_2 \,r^2\,,\nonumber\\
  a(r)^2+b(r)^2&=& Q_0\,,\nonumber\\
  d^2(r)+a(r)^2 b(r)^2&=&\Delta_0 \label{ansmetrcomp}\,.
  \eea
  Moreover, the equations of motion fix  the
  constant parameters $b_0, b_2, c_0, Q_0$, leaving the values of $b_1$, $\Delta_0$ free (although their sizes  must be contained within certain intervals). Notice that
   in General Relativity, diffeomorphism invariance allows one to choose the function $c(r)$ to be
  $c(r) \,=\,r$, so that $c_0=1$. In
   this theory of massive gravity, after having fixed
  the gauge, this choice is no longer possible and the equations
  of motion determine $c_0$. One finds
%    In order to do this, one observes
  %that the metric Ansatz (\ref{genmetr})
  %leads to the following identity between
 % components of the
 % Einstein tensor:  $b(r)^2 G_{rr}+a(r)^2 G_{tt}\,=\,0$.
 % This combination on the energy momentum tensor provides the following
 % value for
 %$c_0$,
 \be\label{solb0}
c_0 \,=\,
 \frac{1 + 6 \a_3 + 12 \a_4 \pm\sqrt{1 + 3 \a_3 + 9 \a_3^2 - 12
\a_4}}{3 (1+3 \a_3 +4 \a_4)},
\ee
for non $\a_3\neq -4\a_4$, and
\be\label{solb0_a3a4}
c_0 =\frac{2}{3}\left(
 \frac{1-12 \a_4}{1-8 \a_4}
 \right)\nonumber,\\
\ee
for $\a_3= -4\a_4$, which in particular includes the case $\a_3=\a_4$=0.
% The upper branch
% generalizes the result of \cite{us}, while the lower
% branch is specifically associated
%  with theories in which $\alpha_3$ and/or $\alpha_4$ are non-vanishing.
    After plugging the metric components
(\ref{ansmetrcomp}) in the remaining Einstein equations,
one can find the values for the other parameters. The corresponding
general
 expressions are quite lengthy, and for this reason we relegate them
 to Appendix \ref{AppA}.
 As a concrete, simple
 example, in the main text we
  work out the special case
 $\a_3=-4 \,\a_4$, where the parameters are
 \bea
 {c_0} =\frac{2}{3}\left(
 \frac{1-12 \a_4}{1-8 \a_4}
 \right)\,,\qquad && \qquad
 b_0=\frac{\Delta_0}{c_0^2}\nonumber,\\
 b_2=\frac{m^2\,\Delta_0}{4\left(12 \a_4-1\right)}
 \,,\qquad&&\qquad
 Q_0=
 \frac{
 16 (1 - 12 \a_4)^4 + 81 (1 - 8 \a_4)^4 \Delta_0}{36 \left[
 1 + 4 \a_4 (-5 + 24 \a_4)\right]^2}.
 \eea
The previous solution is valid for $\alpha_4$ in the ranges
$\alpha_4< 1/12$ and $\alpha_4 > 1/8$. %Notice that the case $\alpha_4
%= 1/12$ corresponds exactly to the Lagrangian (\ref{nambueq}), discussed in
% Section \ref{gensols}.
  We find that
  $b_1$ and $\Delta_0$ are arbitrary; this vacuum solution is then
  characterized by two integration constants.
The resulting metric coefficients
 can be rewritten in the following, easier-to-handle form:
\bea
a(r)^2&=&\frac94 \,\Delta_0\,
\left(\frac{1-8 \a_4}{1-12 \a_4}\right)^2\,\left[ p(r) +\gamma+1\right]
\hskip0.5cm,\hskip0.5cm c(r)\,=\,\frac{2}{3}\,
\left(\frac{1-12 \a_4}{1-8 \a_4}\right)\,r\nonumber\\\label{solrew}
b(r)^2&=&\frac94 \,\Delta_0\,
\left(\frac{1-8 \a_4}{1-12 \a_4}\right)^2\,\left[1- p(r) \right]
\hskip0.5cm,\hskip0.5cm \\d(r)&=&
\frac{9 \Delta_0}{4} \,\left(\frac{1-8\a_4}{1-12\a_4}\right)^2\,
\sqrt{p(r) \left(p(r)+\gamma\right)}\nonumber
\eea
with ($\mu=-b_1/b_0$)
\be\label{defpb}
p(r)\,\equiv\,\frac{\mu}{r}+\frac{\left(1-12 \a_4\right)\,m^2\,r^2}{9\,(1-8 \a_4)^2}
\,\hskip0.8cm,\hskip0.8cm \gamma\,\equiv\,\frac{16}{81 \Delta_0}\,
 \left(
\frac{1-12 \a_4}{1-8 \a_4}\right)^4
-1\,.\ee

In order to have a consistent solution, we must demand
 that the argument of the square root appearing in the expression
 for $d(r)$, Eq.~(\ref{solrew}),  is positive. A  sufficient condition to ensure
 this is that $\mu \ge 0$, and
 \be
 0\, <\,
  \sqrt{\Delta_0}\,<\,\frac{4}{9}\,
  \left(
\frac{1-12 \a_4}{1-8 \a_4}\right)^2\,.
 \ee

\smallskip

The metric  might be rewritten in a more transparent diagonal
form, by means
of a coordinate transformation. In absence of diffeomorphism invariance, any   coordinate transformation
of time $t$ forces us to leave the unitary gauge, and to  switch  on a non-trivial profile
% by
%a
for the St\"uckelberg field
%
%coordinate is not permitted,  since until this
%point we have adopted the
%unitary gauge. Therefore, we  now renounce to this gauge choice,
%and allow for a non-zero vector
$\pi^\mu$ of the form $\pi^\mu\,=\,\left( \pi_0(r),0 ,0 ,0\right)$. One finds that
then the metric can be rewritten in a diagonal form, as
\be\label{newfmetr}
d s^2\,=\,-b(r)^2 d t^2+\tilde{a}(r)^2 d r^2+c(r)^2\,d \Omega^2,
\ee
while the equations of motion for the fields involved are solved by
\be
\tilde{a}(r)^2\,=\,\frac{4}{9}\,
\left(\frac{1-12 \a_4}{1-8 \a_4}\right)^2\frac{1}{1-p(r)}
\hskip0.8cm,\hskip0.8cm \pi_0'(r)\,=\,-\frac{\sqrt{p(r) (p(r)+\gamma)}}{1-p(r)},\label{solffields}
\ee
with $b(r)$, $c(r)$ and $p(r)$ being the same as in Eq.~(\ref{solrew}).
If one then makes  a further time
rescaling \be t\to \frac{ 4
\left(1-12 \a_4\right)^2 }{
9 \Delta_0^{1/2}
\left(
1-8 \a_4\right)^2}\,t\; ,
\label{ftire}\ee
 the resulting metric acquires a manifestly de Sitter-Schwarzschild,
or Anti-de Sitter-Schwarzschild form. The choice between these two possibilities
  depends on whether  $\alpha_4$
is smaller or larger than $1/12$, as can be seen inspecting
the function $p(r)$ in Eq.~(\ref{defpb}).
% by making a time
%rescaling $t\to \left[ 4
%\left(1-12 \a_4\right)^2\right]/\left[
%9 \Delta_0^{1/2}
%\left(
%1-8 \a_4\right)^2\right]\,t
%$.
 On the other hand, we should point out that this time-rescaling
cannot be performed, without further introducing a time dependent
contribution to $\pi_0$. As expected, the metric in Eq.~(\ref{newfmetr})
can also be obtained by making the following transformation
of the time coordinate $d \tilde t\,=\,d t +\pi_0' dr$ to the original
 metric
(\ref{genmetr}). This produces a non-zero time component for
$\pi^\mu$, that does not vanish even in the limit $m\to 0$.

\smallskip

To summarize so far, we found vacuum solutions in this theory that
are  asymptotically de Sitter or Anti-de Sitter, depending
on the choice of the parameters.
 %The value of the
%curvature is ruled by the mass of the graviton: in the case
%of de Sitter space, this fact can explain the smallness of the
%observed cosmological constant in terms of the smallness
%of the graviton mass.
%Let us point out that it is also possible to include
%a bare cosmological constant,
%$\Lambda$, as in the original Lagrangian (\ref{genlag}). Our solutions
%to the Einstein equations, with our metric Ansatz, remain
%formally identical. The only difference is that the function $p(r)$ in Eq.~(\ref{defpb})
% becomes
 %\be\label{pforlambda}
 %p(r)\,=\, \frac{\mu}{r}+\frac{\left(1-12 \a_4\right)}{9\,(1-8 \a_4)^2}
 %\,\left[m^2+\frac43 \left(1-12 \a_4\right)\Lambda
 %\right]\,r^2.
% \ee
%Notice that the additional integration constant $\Delta_0$
%can not be used to 'compensate' the contribution
%of the  bare cosmological
%constant $\Lambda$ via a self-tuning mechanism, since $\Delta_0$ does not explicitly appear in the
%previous formula.

Figure \ref{fullsolutions} shows the allowed parameters $\a_3$ and $\a_4$ for the existence of these asymptotically dS or AdS solutions. Further solutions and studies on black holes in this massive gravity theory can be found in \cite{sbh}.

\begin{figure}[htp!]
\begin{center}
\includegraphics[width=6cm]{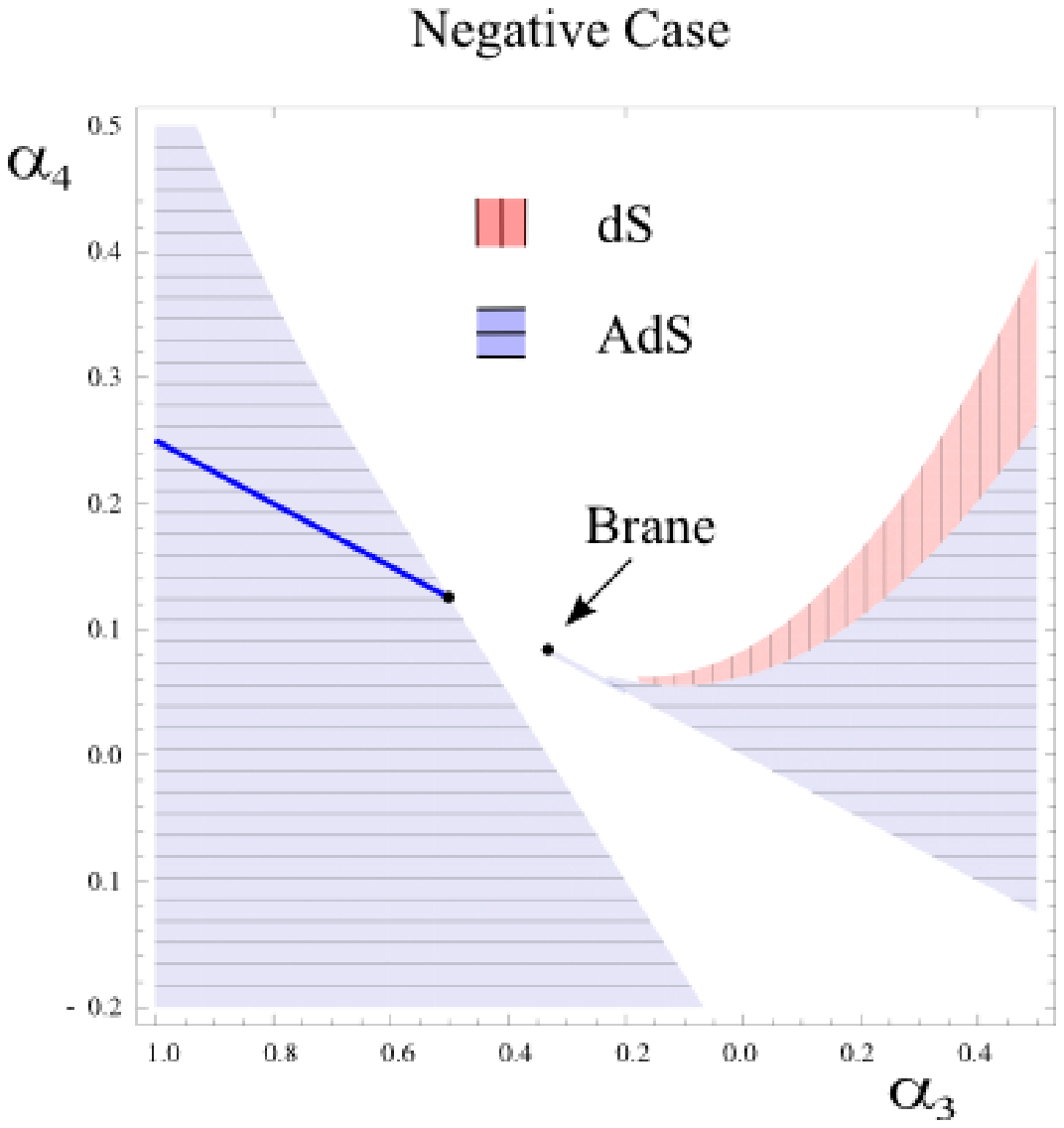}\hspace{1cm}
\includegraphics[width=6cm]{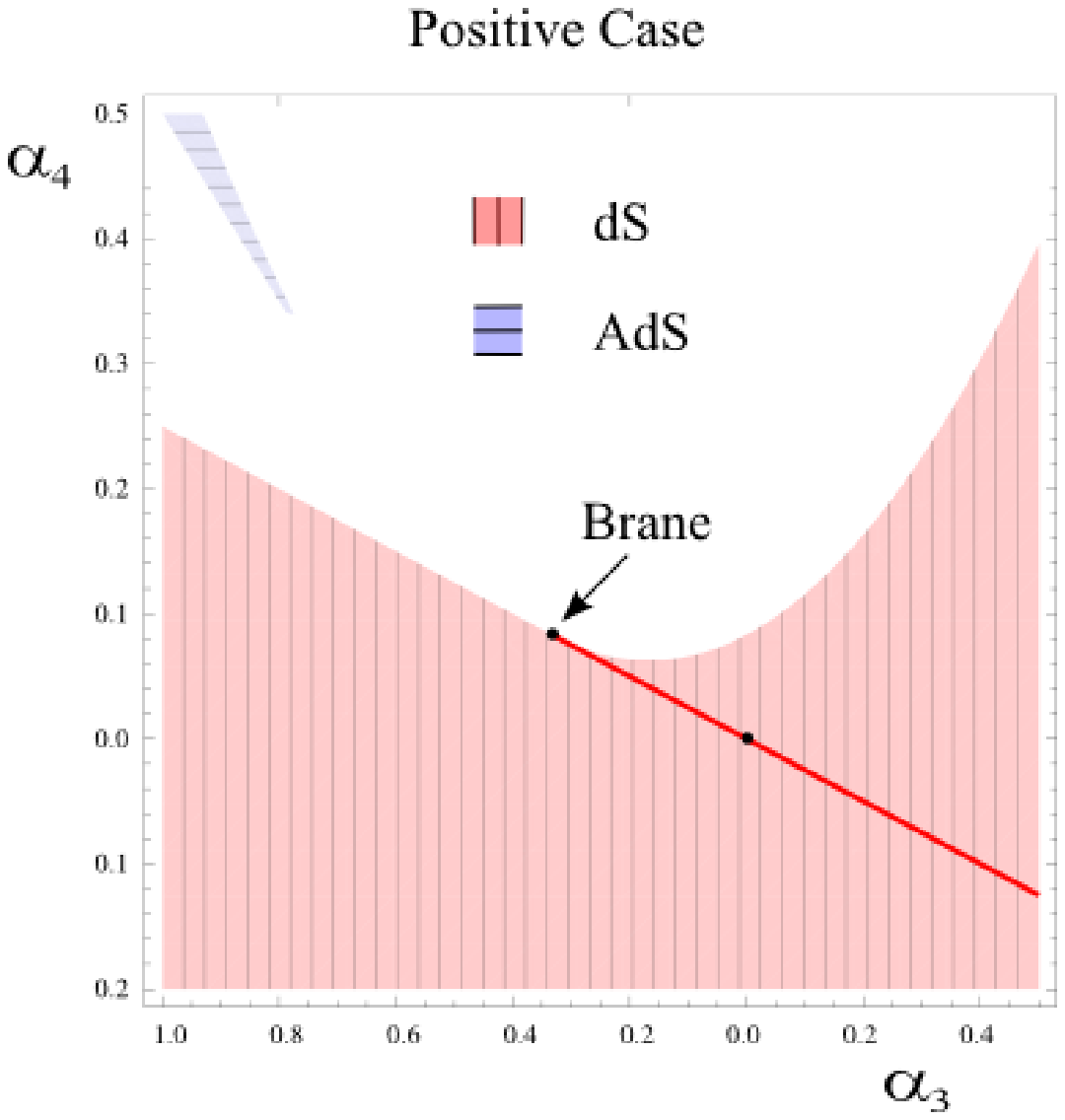}
\caption{Allowed region of parameter space where solutions, which asymptote dS (vertical lines in light or red color) or AdS (horizontal lines in dark or blue color), exist. Left (right) plot is for the negative (positive) branch of Eq.~(\ref{solb0}). The brane solution is given by the special choice of for $\alpha_3=-1/3$ and $\alpha_4=1/12$ \cite{Koyama:solutions}, and the solid lines correspond to the line $\a_3=-4\a_4$. We have set $\Lambda=0$.}
\label{fullsolutions}
\end{center}
\end{figure}

\section{Cosmological Acceleration}\label{cosmosec}

\subsection{Self-accelerating solution}
One of the interesting features of massive gravity is {\it self-acceleration}. The self-accelerating solution was originally found in the DGP braneworld model where the acceleration of the Universe can be realised without introducing the cosmological constant \cite{DGPcosmo}. However the self-accelerating solution in the DGP model suffers from a ghost instability \cite{DGPghost}. 
%{\bf KK I added this paragraph} 

The first complete self-accelerating solution in the $\Lambda_3$ massive gravity theory (\ref{genlag}) was reported in \cite{Koyama:solutions, Koyama:prl} (the self-accelerating solution in the decoupling limit was first obtained in \cite{deRham:2010tw}). This configuration describes an accelerating cosmological universe in the vacuum, in which the rate of acceleration is controlled by the size of the graviton mass.
The solution is a coordinate transformation of the exact solution (\ref{newfmetr})-(\ref{solffields}), after
 having performed the time rescaling (\ref{ftire})
 (one can use (\ref{boapp})-(\ref{papp}) for more general values of $\a_3$ and $\a_4$). Let us review how to construct it for the simple case of $\a_3=-4\a_4$. In the case of the asymptotically de Sitter solution (\ref{newfmetr})-(\ref{solffields}) with $\alpha_4<1/12$ and $\mu=0$,
the metric can also be written in a time dependent form, at the price of switching on additional components of $\pi^\mu$.  After dubbing
$$\tilde m^2\equiv
\frac{m^2}{(1-12 \a_4)}
 %\,\left[m^2+\frac43 \left(1-12 \a_4\right)\Lambda
 %\right],
$$
we can make the following coordinate transformation
$t\,=\,F_t(\tau, \rho)$ and $r\,=\,F_t(\tau, \rho)$ with
\bea
F_t(\tau, \rho)&=&\frac{4}{3 \,\Delta_0^{1/2} \,\tilde{m}}\,
\Big(\frac{1-12 \alpha_4}{1- 8 \alpha_4} \Big)
\,{\rm arctanh}\,\left(
\frac{\sinh{\left(\frac{\tilde m \tau}{2}\right)}+\frac{\tilde{m}^2 \rho^2}{8} e^{
\tilde{m} \tau/2}}{\cosh{\left(\frac{\tilde m \tau}{2}\right)}-\frac{\tilde{m}^2 \rho^2}{8} e^{
\tilde{m} \tau/2}}
\right),
\\
F_r(\tau, \rho)&=&\frac32 \Big(\frac{1-8 \alpha_4}{1-12 \alpha_4}\Big)\rho \, e^{
\tilde{m} \tau/2}.
\eea
Then, the metric becomes that of flat slicing of de Sitter
\be\label{solmetric}
  d s^2\,=\,
- d \tau^2+ e^{{\tilde m} \tau}\,\left( d \rho^2+\rho^2  d \Omega^2\right),
\ee
where the Hubble parameter is given by
\be\label{Hubble}
H\,=\,\frac{\tilde m}{2}\,=\,\frac{m}{2 (1-12 \a_4)^{\frac12}}
%\,\left[m^2+\frac43 \left(1-12 \a_4\right)\Lambda
%\right]^\frac12.
\ee
The St\"uckelberg fields $\pi^\mu$ are now given by
\be\label{solpis}
\pi^\mu\,=\,\left( \,\pi_0+\tau-F_t, \, \rho-\,F_r,\,0,\,0 \right).\
\ee
Interestingly, the value of the Hubble parameter is ruled by the mass of the graviton:
  we have a self-accelerating solution, in which the smallness of the 
   acceleration rate
%  observed cosmological constant 
 can be associated with the smallness of the graviton mass.

%{\bf KK  I edited cosmological solutions part below:}

\bigskip 
So far, we used the general static solutions of the previous section to determine time-dependent self-accelerating configurations 
via suitable coordinate transformations. However, one can also follow another approach, and try to directly find time-dependent, self-accelerating 
configurations in the theory of $\Lambda_3$ massive gravity without relying on the unitary gauge. The hope, following this second route,  is to determine 
 additional cosmological configurations for this  theory. Starting from the Ansatz we wrote in eq. (\ref{metricatz}), 
more
 general cosmological solutions can be obtained by focussing on the Ansatz  $d(r,t)=0$ and $c(r,t)\,=\,r\, a(r,t)$ in (\ref{metricatz}) so that the metric becomes
\be
ds^2 = - b(t,r)^2 dt^2 + a(t,r)^2 (dr^2 + r^2 d \Omega^2).
\ee
Wyman et al \cite{wyman} showed that self-accelerating configurations are characterised by the following profile for the function
 $g(r,t)$ 
  characterizing the St\"uckelberg field  $\phi^i$ (see eq (\ref{stuckatz}))

\be
g(t,r)= c_0^{-1} a(t,r) r,
\label{selfacc-cond}
\ee
where $c_0$ is given by (\ref{solb0}).
The equation of motion for $g$ evaluated on the solution (\ref{selfacc-cond}) provides a constraint on the function  $f$
characterizing the Struckelberg field  $\phi^0$
\be
\sqrt{\tX} P_1' =\left( 2 c_0 P_2-P_2' \right) W  - P_0',
\label{eqn:Xfeom}
\ee
where the $P_n$ functions
\begin{eqnarray}
P_0(x) &=& - 12 - 2 x(x-6) - 12(x-1)(x-2)\alpha_3 -24(x-1)^2\alpha_4, \nonumber\\
P_1(x) &=& 2 (3 -2 x)  +  6(x-1)(x-3)\alpha_3 +   24(x-1)^2 \alpha_4, \nonumber\\
P_2(x) &=& -2 + 12 (x-1) \alpha_3 - 24(x-1)^2 \alpha_4.
\end{eqnarray}
are evaluated at $x=c_0^{-1}$, $P_n'(x)\equiv dP_n/dx$. Moreover,
  
\be
\tX = \Bigl(\frac{\dot{f}}{b}+\mu\frac{g'}{\ta}
\Bigr)^2-\Bigl(\frac{\dot{g}}{b}+\mu\frac{f'}{\ta }\Bigr)^2, \quad
W = \frac{\mu}{ab} ( \dot f g' - \dot g f' ),
\ee

\noindent
and $\mu={\rm sgn}(\dot f g' - \dot g f')$
where primes denote derivatives with respect to $r$ and overdots with respect to $t$.
Using these equations, it is possible to show that the Einstein equations are given by
\be
G^{\mu}_{\nu} = - 3 H^2 \delta^{\mu}_{\nu}, \quad H^2=\frac{1 + 3 \alpha_3 \pm 2 \alpha_5}{ 3(1 + 3 \alpha_3 \pm \alpha_5)^2} m^2,
\label{Einstein}
\ee
where \be \alpha_5^2\equiv1+3\alpha_3+9\alpha_3^2-12\alpha_4\,.\label{defda5}\ee
 Note that there are two branches of solutions. This
 approach 
 leads to self-accelerating solutions where the Hubble parameter is determined by the mass of the graviton.

There are many possible solutions for the function $f(t,r)$ for given solutions of metric satisfying the Einstein equations (\ref{Einstein}). For example, in the simple case of $\alpha_3=\alpha_4=0$, the configuration (\ref{solmetric}) and (\ref{solpis}) is given by
\be
b=1, \quad a=e^{m t/2}, \quad
f(t,r) = -\frac{3}{2} e^{m t/2} r + \frac{3}{m} \left(
{\rm arctanh} \left(\frac{1}{2} e^{m t/2} m r\right) +
{\rm arctanh} \left( \frac{4-e^{m t} (4 + m^2 r^2)}{-4 + e^{m t} (-4 + m^2 r^2)}\right)
\right).
\ee
Notice that this self-accelerating solution has a flat FRWL metric, however, one can also write it as an open or closed FRWL spacetime, with the price of changing the St\"uckelberg fields accordingly. In all the FRWL frames, the St\"uckelberg fields are inhomogeneous. In fact it was suggested that there was no FRW solution that keeps the
FRW symmetry for the fiducial metric $\Sigma^{\mu}_{\nu}$ \cite{D'Amico:2011jj}. However, Gumrukcuoglu et al found a special self-accelerating solution which represents an open universe where the fiducial metric $\Sigma_{\mu\nu}$ respects the FRW symmetries of the physical metric $g_{\mu\nu}$ \cite{Gumrukcuoglu:2011ew}. Their solution is given by
\be
b=1, \quad a=\frac{a_0(t)}{1 - (m r)^2/16}, \quad f(t,r)=\frac{3}{m} a_0(t) \frac{1+ (m r)^2/16}{1- (m r)^2/16}, \quad a_0(t)=\sinh (m t/2).
\ee
For this solution the fiducial metric preserves the FRW symmetry, 
$\Sigma^{\mu}_{\nu}= (9/4) {\rm diag}((2\dot{a}_0/m)^2, 1,1,1)$.
The behaviour of perturbations around this particular self-accelerating solution is very different from the other solutions that break the FRW symmetry for the fiducial metric $\Sigma^{\mu}_{\nu}$. At the linear order, scalar and vector perturbations have no kinetic terms, hence  they are strongly coupled \cite{Gumrukcuoglu:2011zh}, which leads to non-linear instabilities \cite{DeFelice:2012mx, DeFelice:2013awa}. The absence of the scalar kinetic term originates from the special choice of the solution for $f$ \cite{Wyman2} that retains the FRW symmetry for the fiducial metric. This leads to an enhanced symmetry that  eliminates the scalar perturbations \cite{new}.
In the rest of this review, we do not consider this class of self-accelerating solution and consider the case where the FRW symmetry is broken for the fiducial metric $\Sigma^{\mu}_{\nu}$. However, we emphasise that the physical metric still retains the FRW symmetry in these solutions.

Wyman et al \cite{wyman} also showed that the ordinary Friedmann equation is obtained even if we add ordinary matter energy density $\rho_m(t)$. The matter only sees the effect of the mass terms as a cosmological constant with no direct coupling to the scalar fields on the exact solution. Cosmological solutions in massive gravity and its extension including de a Sitter fiducial metric and bigravity can be found in \cite{cosmorefs}.

\subsection{Decoupling limit solutions and their instability}

Once we determined  self-accelerating, de Sitter solutions in this model, 
  it is crucial to study their stability:
  this
  is
   the subject of this section.
%   what we are going to do 
 %in this section. 
 In order to make the analysis manageable, we focus on 
 a convenient limit of Lagrangian (\ref{genlag}) which captures most of the dynamics of the helicity-0 and helicity-1 mode, but keeps the linear behaviour of the helicity-2 (tensor)   mode \cite{ArkaniHamed:2002sp}. %Moreover, in this limit the theory still describes interactions between the scalar and the vectors, and also the scalar with the tensors.
The limit, called the decoupling limit, is defined as
\be\label{declimit}
m\to 0\,,\hskip1cm M_{Pl}\to \infty\,,\hskip1cm \Lambda_3\,\equiv\,m^2M_{Pl}=\,{\text{fixed}}\,,
\ee
In order to obtain canonically normalized kinetic terms for the helicity 2 and helicity 1 modes, together with the relevant couplings for the helicity 0 modes, when this limit is taken one needs to canonically normalise the fields in the following way
\be\label{cannorm}
h_{\mu\nu}\,\to\,M_{Pl}\, h_{\mu\nu}
\,, \hskip 0.5cm A_{\mu}\,\to\,m M_{Pl}\, A_\mu
\,, \hskip 0.5cm \pi\,\to\,m^2 M_{Pl}\, \pi,
%\,, \hskip 0.5cm \Lambda\,\to\,M_{pl}\, \lambda_0,
\ee
where we have split the St\"uckelberg fields $\pi^\mu$ into a scalar component $\pi$ and a divergenceless vector $A^\mu$  in the usual way, namely
\be\label{vecscal}
\pi^\mu=\eta^{\mu\nu}(\partial_\nu \pi +A_\nu).\ee

In order to take the decoupling limit (\ref{declimit}) of the self-accelerating solutions defined by (\ref{selfacc-cond}), the solution has to be in the conformally flat frame, which is defined as follows
\be
a(t,r)=b(t,r)=\frac{c(t,r)}{r}=(1+H^2(r^2-t^2)/4)^{-1}.
\ee
In this frame, all the known self-accelerating solutions 
%(including the one presented here and that of Gumrukcuoglu et al) 
lead  to the same decoupling limit solution for the St\"uckelberg fields, namely
\bea\label{pisdec}
\pi^0&=&\left(\frac{1+3 \alpha_3+\alpha_5}{(2+3
   \alpha_3+\alpha_5)\sqrt{\Delta_0} }-1\right)t
   -\frac{1}{2}\sqrt{\frac{(1+3 \alpha_3+2 \alpha_5) \left[(1+3
   \alpha_3+\alpha_5)^4-(2+3 \alpha_3+\alpha_5)^4\Delta_0 \right]}{
   3\Delta_0 (1+3 \alpha_3+\alpha_5)^4 (2+3
   \alpha_3+\alpha_5)^2}}m r^2 +\mathcal{O}(m^2)\nonumber \\
\pi^r&=&-\frac{1}{1+3 \alpha_3+\alpha_5}r+\mathcal{O}(m^2),
\eea
with $\alpha_5$ given by eq. (\ref{defda5}). 
If we split $\pi^\mu$ into a scalar and vector piece as in (\ref{vecscal}), canonically normalise the fields as in (\ref{cannorm}), and take the decoupling limit (\ref{declimit}), one gets \cite{Koyama:vect1, Koyama:vect2}
\bea
h_{\mu\nu}&=&-\frac{1}{2}\Lambda_3 H^2(r^2 - t^2)\eta_{\mu\nu},\nonumber\\
\pi&=&-\frac{1}{2}\left(1-\frac{1}{c_0}\sqrt{1+\frac{3  (1+3 \alpha_3+\alpha_5)^4Q_0^2}{(2+3
   \alpha_3+\alpha_5)^2 (1+3 \alpha_3+2 \alpha_5)%+(1+3 \alpha_3+\alpha_5)^2\Lambda]
   }}\right)\Lambda_3 t^2 -\frac{c_0-1}{2\, c_0}\Lambda_3 r^2,\nonumber \\
A_0&=& -\frac{Q_0}{2}\Lambda_3 r^2,
\eea
where $c_0$ is given by (\ref{solb0}) (or (\ref{solb0_a3a4}) for $\a_3=-4\a_4$), and the Hubble parameter $H$ by
\be
H^2=\frac{(1 + 3 \alpha_3 \pm 2 \alpha_5) m^2 %+(1 + 3 \alpha_3 + \alpha_5)^2\Lambda
}{ 3(1 + 3 \alpha_3 \pm \alpha_5)^2},
\ee
%{\bf KK I corrected this, but please check} 
 which is the generalisation of (\ref{Hubble}) for arbitrary $\a_3$ and $\a_4$. Moreover, there is a relation  between %, the previously used, integration constant 
 $\Delta_0$ and $Q_0$ 
\be\label{delta0}
\Delta_0=\frac{(1+3 \alpha_3+\alpha_5)^4 %[
(1+3 \alpha_3+2 \alpha_5)%+(1+3 \alpha_3+\alpha_5)^2\Lambda
%]
}{3
   Q_0^2 (2+3 \alpha_3+\alpha_5)^2 (1+3
   \alpha_3+\alpha_5)^4+(2+3 \alpha_3+\alpha_5)^4 %[
   (1+3 \alpha_3+2 \alpha_5)%+(1+3 \alpha_3+\alpha_5)^2\Lambda]
   },
\ee
and in the case of AdS, there is an extra bound given by
\be
Q_0^2<\frac{(2 + 3 \alpha_3 + \alpha_5)^2%\Big|
(1+3 \alpha_3+2 \alpha_5)%+(1+3 \alpha_3+\alpha_5)^2\Lambda\Big|
 }{3(1 + 3 \alpha_3 + \alpha_5)^4}.
\ee
%The vector mode has only being seen in the solution (\ref{solmetric})-(\ref{solpis}), but not in the Gumrukcuoglu et al. It may be possible to construct a solution of the full theory based on Gumrukcuoglu et al which presents some vector charge, but it is beyond this work.
If one take the vector charge to zero $Q_0=0$ (or equivalently if $\Delta_0=(1+3
\alpha_3+\alpha_5)^4/(2+3 \alpha_3+\alpha_5)^4$), these solutions can be written in a simpler covariant way
\be\label{back_sols}
h_{\mu\nu}=-\frac{1}{2}\Lambda_3 H^2(x^\mu x_\mu)\eta_{\mu\nu}\qquad \qquad \pi=\frac{c_0-1}{2\,c_0}\Lambda_3 x^\mu x_\mu, \qquad \qquad A^\mu=0.
\ee
Therefore, corrections of order $m^2$ in (\ref{pisdec}), which do not show in the decoupling limit, are the main differences among solutions in the full theory. These solutions in the decoupling theory were also found in \cite{deRham:2010tw}.

\smallskip
 
 In this decoupling limit, the
  structure of the Lagrangian becomes much   simpler, 
and  the various self-accelerating configurations become the same.
 For these reasons,
   it is particularly   convenient 
to study the dynamics of  perturbations
 in this limit. If problems or instabilities arise in this limit, then are they unavoidably present also in the full theory outside 
 the decoupling regime. 
 % for a large class of self-accelerating solutions in a systematic way.
 Interestingly,
 it has been shown in \cite{deRham:2010tw} that, in the decoupling limit,  the coupling
  between the scalar mode $\pi$ and the trace $T$ of the energy momentum tensor vanishes around these self-accelerating
  configurations: hence, the coupling to matter is the same as in GR with no need to implement a Vainshtein mechanism.
   However, we have shown
    in \cite{Koyama:vect1, Koyama:vect2}  that all these backgrounds present instabilities in the vector sector.  We present here the main results concerning these instabilities,
focussing on the case of $\a_3=\a_4=0$. A generalisation to arbitrary values is straighforward and can be found in \cite{Koyama:vect1, Koyama:vect2}. 
%A different argument is developed in \cite{DeFelice:2012mx}.
 We start by considering perturbations of the fields $h_{\mu\nu}$, $A^\mu$ and $\pi$ which only depend on time and radial
 component.  Namely
\be\label{perts}
h_{\mu\nu}=h^0_{\mu\nu}+\hat{h}_{\mu\nu}, \qquad \qquad A^\mu=A^\mu_0+\hat{A}^\mu, \qquad \qquad \pi=\pi_0+\hat{\pi},
\ee
where the background quantities (those with an index $0$) are given by the self-accelerating solution (\ref{back_sols}).
The Lagrangian for the tensor and scalar perturbations (without further truncations) reads
% {\bf GN I had evaluated $c_0=3/2$ and $H^2=1/4+\lambda_0$ everywhere}
\bea\label{lagdec2}
{\cal L}_{h_{\mu\nu},\;\pi}&=& -\frac12 \,\hat{h}^{\mu \nu} {\cal E}_{\mu \nu}^{\alpha \beta} \hat{h}_{\alpha\beta}
%- 3 \left[  \lambda_0 -   H^2+\frac{1}{4}\right]\,  \hat{h}
%-12   H^2 \frac{(3c_0-2)}{c_0} \hat{ \pi}
+\hat{h}^{\mu\nu} {X}_{\mu\nu}-6  H^2\,\hat{\pi} \Box \hat{\pi}\,,
\eea
where (in units in which $\Lambda_3=1$, that we adopt from now on) $H^2=\frac{1}{4}%+\lambda_0
$ and $\hat{X}_{\mu\nu}$ is given by
\bea
\hat{X}_{\mu\nu}&=&
 \Big[ %(3-2 c_0)\eta_{\mu\nu} \hat \Pi
 %-(3-2 c_0) \hat \Pi_{\mu\nu}
 \hat \Pi_\mu^\lambda \hat  \Pi_{\lambda \nu}-\hat \Pi \hat \Pi_{\mu\nu}+\frac12\eta_{\mu\nu}\left( \hat \Pi^2-\hat \Pi_{\mu}^\nu \hat \Pi_\nu^\mu\right)
 \Big]\,.
\eea
We can use the following field redefinition to decouple the helicity 2 from the helicity 0 field:
\be
\hat{h}_{\mu\nu}\,\to\,\hat{h}_{\mu\nu}- \partial_{\mu} \hat{\pi} \partial_\nu \hat{\pi}\,.
\ee
Then the kinetic terms for tensor and scalar are diagonalized resulting, up to total derivatives, in
\bea
{\cal L}_{h_{\mu\nu},\;\pi}&=&-\frac12  \hat{h}_{\mu\nu} {\cal E}^{\mu\nu\alpha\beta}  \hat{h}_{\alpha\beta}
%- 3\,\left[  \lambda_0 -   H^2+\frac{1}{4}\right]\, \left( \hat{h}
%+4 (3-2 c_0) \hat{\pi}
%+\hat{\pi} \Box \hat{\pi}\right)
%\nonumber\\
%&&%-12    H^2 (3-2 c_0) \hat{ \pi}
%+\left[\frac32 \left(3-2 c_0\right)^2
-6  H^2%\right]
\hat{\pi} \Box \hat{\pi}%-\frac32 \left(3-2 c_0\right) \left(\partial \hat \pi \right)^2 \Box \hat \pi
 %\nonumber\\
%&&
-\frac12 \left(\partial \hat \pi \right)^2 \left[\left(\Box  \hat \pi \right)^2
- \left( \partial_\mu \partial_\nu \hat  \pi \partial^\mu \partial^\nu\hat \pi\right) \right]\label{lagdec3}
\,.\eea
% Let us emphasize a crucial feature of the previous Lagrangian: the decoupling limit and diagonalization procedures provide a contribution to the kinetic term of the scalar fluctuations, which includes a term proportional to the curvature $H^2$ of space-time. This term is inherited from a third order interaction between tensor and scalar, which produces a second order interaction once the background is included from eq.~(\ref{perts}).
% By considering perturbations around a curved space, the scalar component of the massive graviton acquires a kinetic term which depends on
% the space-time curvature. This was already noticed in \cite{ArkaniHamed:2002sp}, in the context of the Fierz-Pauli theory.
Let us emphasize that the previous Lagrangian contains terms which are quadratic on $\hat{h}_{\mu\nu}$, but higher orders in the scalar field $\hat{\pi}$. The scalar field terms are the so called Galileon combinations. On the contrary, as mentioned before, the vector piece has an infinite number of interactions \cite{Koyama:vect1, Koyama:vect2}. %so we need to truncate it to a given order.
For our purposes, it is enough to stop at fourth order in the fields, resulting in the Lagrangian
 \bea\label{fvlags}
{\cal L}_{A_\mu}&=&\frac{1}{18} \Big\{ %(3-2 c_0)\,F^{\mu\nu}F_{\mu\nu}
 3\hat \Pi \,F^{\mu\nu}F_{\mu\nu}%+\frac{(3-4 c_0)}{ c_0}
 -6\,\left( \hat{\Pi}^{\alpha}_{\ \mu} F^{\mu\nu}F_{\nu\alpha}\right)\nonumber
 \\&&%+\frac{\left(3-4 c_0\right)}{2\, c_0^2}
 -2\left[ \hat{\Pi}^\mu_{\ \nu}\hat{\Pi}^\nu_{\ \alpha} F^{\alpha\beta}F_{\beta\mu}+\hat{\Pi}^\mu_{\ \nu}F^{\nu\alpha}\hat{\Pi}^{\ \beta}_{\alpha} F_{\beta\mu}
-  \hat \Pi \, (\hat{\Pi}^{\alpha}_{\ \mu} F^{\mu\nu}F_{\nu\alpha})\right]
  \Big\}
  \,+\,\dots
 \eea
 where $F_{\mu\nu}=\partial_\mu A_\nu-\partial_\nu A_\mu$.
 As mentioned above,
the kinetic terms for the vectors vanish; however, vectors become dynamical by coupling them with the scalar at third or higher order in fluctuations (this was already pointed out in \cite{deRham:2010tw,Koyama:vect1,D'Amico:2012pi}). Nevertheless, one should worry about
higher derivatives in the equations of motion, since the previous Lagrangian contains contributions with two time derivatives in the scalar field $\pi$. For systems coupling scalars with vectors, it is possible to find the combination  that ensures that the equations of motion do not contain at all terms containing more than two time derivatives. It is a generalization of Galileon combinations which was explored in \cite{Deffayet:2010zh} and dubbed $p$-form Galileons. Up to fourth order in perturbations, the correct combination (without including higher derivatives in $A_\mu$)
%{\bf GN I believe this is the case, since higher derivatives in A may give a different answer, but I couldn't probe that was the case, since i think i missed possible terms} {\bf GT not clear to me, but fine to leave as you wrote})
  is
 \be
{\cal L}^{p-form}\,=\,a_0\,\left[ \tr \Pi\,\tr F^2 -2 \,\tr \Pi F^2 \right] +b_0\,\left\{
\tr F^2\,\left[\tr \Pi^2 -\left( \tr \Pi\right)^2\right]-4 \tr \Pi\, \tr \Pi F^2 + 4 \tr \Pi^2 F^2 +2  \tr \Pi F \Pi F
%Tr[F2] (Tr[P]^2 - Tr[P^2]) - 4 Tr[P] Tr[P.F2] + 4 Tr[P.P.F2] +
% 2 Tr[P.F1.P.F1]
\right\}\label{pfgal}
\ee
where $a_0$ and $b_0$ are arbitrary coefficients, and we have used the $\tr$ notation to simplify the index structure. The above third order action with the aforementioned properties
was presented in \cite{Deffayet:2010zh}, while the fourth order one is as far as we know new.
 Comparing (\ref{fvlags}) with (\ref{pfgal}) we notice that
 while the third order action has the correct structure to avoid higher order time derivatives in the equation of motion,
 the fourth order Lagrangian does not seem to satisfy this requirement. However, a suitable field redefinition allows
 to remove the fourth order term from the third order contribution, leaving a healthy Lagrangian without higher derivative equations of motion (in agreement with the ghost-free statement of the theory).

\bigskip

On the other hand, although our scalar-vector Lagrangian (\ref{fvlags}) does not lead to a propagation of a sixth ghost mode, it does
generally  lead to a ghost-like instability around self-accelerating configurations, in which the ghost is one  of the available vector modes. It has be shown in \cite{Koyama:vect1} that, when turning on a non-trivial profile for the background vector field,
 the corresponding Lagrangian for perturbations around the resulting configuration  acquires kinetic terms for the vector
 with the wrong sign. % For completeness, we include a  proof of this fact in Appendix
%\ref{app-vec}, based on the Lagrangian (\ref{fvlags}) expanded up to third order in perturbations.  In the remaining
  Here, following \cite{Koyama:vect2},  we instead directly point out the instability by analyzing the Hamiltonian associated with the Lagrangian
   obtained by combining the third order Lagrangian contained in (\ref{fvlags}) with the scalar kinetic term:
\be\label{thirdoa2}
{\cal L}^{third}\,=\,-3  H^2
{\pi} \Box{\pi}-\frac16 \left[ \Box \pi\,F_{\mu\nu} F^{\mu\nu}-2 \partial_{\mu\nu} \pi F^{\mu\rho} F^{\nu}_{\;\;\rho}\right],
\ee
where we have removed the hats over the field to simplify the notation.
We choose, for simplicity, the gauge $A_0=0$, $\partial_i A_i=0$, and by doing a standard $(3+1)$-decomposition, the previous Lagrangian reads
\be
{\cal L}\,=\,-3 H^2 \dot{\pi}^2+\frac13\,\left[2 \dot{\pi} \,\dot{A}_i  \triangle A_i+\triangle \pi \,\dot{A}_i^2-\pi_{,\,ij} \,\dot{A}_i \dot{A}_j\right]+\dots
\ee
where the dots represent the terms without time derivatives, which we do not include since they do not play a role in the present discussion.
The conjugate momenta to $\pi$ and $A_i$ are
\be
\Pi_{\pi}=-6 H^2 \,\left(\dot{\pi}-\frac{1}{9\,H^2} \,\dot{A}_i  \triangle A_i\right)\,,\qquad \qquad
\Pi_{A_i}=\frac{2}{3}\left[ \dot{\pi} \, \triangle A_i+\triangle \pi \,\dot{A}_i -\pi_{,\,ij} \,\dot{A}_j
\right]\label{conm1}\,.
\ee
In order to analyze the associated Hamiltonian, it is convenient to introduce the matrix
${\kappa}_{ij}\,\equiv\,\triangle \pi \,\delta_{i j} -\pi_{,\,ij}$. If $\kappa_{ij}\,=\,0$, then, we can easily invert the relations that define the conjugate momenta, and obtain the following Hamiltonian
\bea
{\cal H}&=&-\frac{\Pi_\pi^2}{12 \,H^2}+\frac{1}{12 H^2}\, \left(
\Pi_\pi+\frac{9\,H^2\,A_i \Pi_{A_i}}{A_i \,\triangle A_i}
\right)^2+\dots\,,\\
&=&\frac{3}{2}\, \Pi_\pi\,\frac{A_i \Pi_{A_i}}{A_i \,\triangle A_i}+
\frac{27\,H^2}{4 }\, \left(
\frac{A_i \Pi_{A_i}}{A_i \,\triangle A_i}
\right)^2+\dots\,,
%+\frac13\,
%\left( \triangle \pi \,\dot{A}_i^2-\pi_{,\,ij} \,\dot{A}_i \dot{A}_j\right)
 \label{haminf2}
%\left[ \left( \dot{A}_i \,\triangle A_i\right)^2+18 H^2 \left( \triangle \pi \,\dot{A}_i^2-\pi_{,\,ij} \,\dot{A}_i \dot{A}_j\right) \right]
\eea
where the dots represent terms without momentum variables.
 The previous Hamiltonian is linear in $\Pi_\pi$; hence it is unbounded from below. Notice that this argument holds
 even in the limit in which $H^2$ vanishes.
%when one independently increase the size of conjugate momenta.
In conclusion, perturbations of the background
self-accelerating solution, along the direction of  scalar fluctuations such that ${\kappa}_{ij}\,=\,0$,
% which are functions of time only (or which have isotropic spatial dependence)
%towards the direction of $\vec{x}-$independent $\pi$ fluctuations
%would
admit unstable directions along which
  the system falls towards regions where the energy is unbounded from below.
Similar conclusions hold for more generic $\kappa_{ij}$. Let us, for example, consider a $\kappa_{ij}$ that is non-vanishing, and invertible. Then, after straightforward manipulations, one can show that the Hamiltonian can be written as
\be
\frac43\,{\cal H}\,=\,-\frac{1}{9 H^2+\Delta A_i \kappa_{ij}^{-1} \Delta A_j}\,\left(\Pi_\pi-  \Delta A_i \kappa_{ij}^{-1}  \Pi_{A_j}\right)^2
+ \Pi_{A_i} \kappa_{ij}^{-1}  \Pi_{A_j}+\dots\,,
\ee
where, again, the dots represent terms without momentum variables.
It is not difficult to see that there are many unstable directions associated with this Hamiltonian.
 For example,  make a choice  for the vector $\Delta A_i$ so that the scalar combination ${\cal C}\,\equiv\,\Delta A_i \kappa_{ij}^{-1} \Delta A_j$ is non-vanishing and has a given sign. For definiteness, the magnitude  of $\Delta A_i$ is chosen such that
the denominator of the first term has the same sign of ${\cal C}$.
  %   this combination is larger, in absolute value, than the $9 H^2$ contribution to the denominator of the first term.
 Accordingly, choose the vector $\Pi_{A_i}$ such that $ \Pi_{A_i} \kappa_{ij}^{-1} \Pi_{A_j} $ has the same sign of ${\cal C}$ (for example, choose it in the same direction of the $\Delta A_i$).  Then, by choosing a suitable magnitude for $\Pi_\pi$,  it is possible to make one of the two terms in the previous Hamiltonian arbitrarily negative -- hence the Hamiltonian is unbounded from below.
Other cases, such as the case in which $\kappa_{ij}^{-1}$ is non-vanishing but not invertible, can be treated in a similar way. Furthermore, while here we focussed on the case $\alpha_3=\alpha_4=0$ it is straightforward to extend this analysis to the more general case, obtaining the same conclusion (see \cite{Koyama:vect2} for details).

%if $\kappa_{ij}^{-1}$ is positive (or negative) definite,
%by choosing appropriately the size of the vectors $\Delta A_i$ the two terms in the previous lagrangian have opposite sign (no matter what the value of $H^2$ is), and one of them renders the energy unbounded from below.
To summarize, one generically expects instabilities around the self-accelerating
solutions discussed so far: there are many directions in the moduli space of fluctuations along
which the energy is unbounded from below, and towards which the
 %No matter how carefully one
%tunes the boundary values for the fluctuations so to avoid instabilities,
 % in absence of symmetries quantum physics  will  detune initial conditions and push the
  system can be driven into
  dangerous regions. On the other hand, to close this section  
   with a positive perspective, it might very well be that suitable deformations of known solutions (or even completely new configurations)
  exist    that, renouncing to  the symmetries imposed on the Ans\"atze considered so far,  do {\it not}  
  present the  problems discussed above.   
        Very recently, a proposal in this direction has been pushed
       forward in \cite{DeFelice:2013awa}, that
       consider  the possibility  of breaking the isotropy of three dimensional spatial slices to find stable configurations.  
   % For example,  trong coupling problems discussed above in the vector sector can be avoided, for example
    %in cases in which 
   % healthy  kinetic terms for the vector 
     %  arise proportional to  a deformation parameter.
       Still much
       work is needed to clarify this subject and analyze phenomenological consequences of these solutions.

\section{Future directions}\label{futuresec}

Massive gravity is a good theoretical laboratory to study modifications of General Relativity with interesting
phenomenological consequences.  Non-linear self-interactions of massive gravity  in proximity of  a source manage 
to mimic the predictions of linearised General Relativity, hence agreeing with 
 solar-system precision measurements. Moreover, massive gravity offers 
 a concrete set-up for studying models of dark energy in modified gravity scenarios.
 Indeed,  
 at large distances gravity is modified with respect to GR, and the theory  
   admits cosmological accelerating solutions in the vacuum  in which the size of acceleration depends on the graviton mass.
  Dark energy models  built in this way have the opportunity to be technically natural in the  't Hooft sense: in the limit of graviton mass
   going to zero one gains a symmetry, by   recovering the full diffeomorphism invariance of GR.
   Consequently, any corrections to the size of dark energy must be proportional
   to the (tiny) graviton mass itself.

   Hence, non-linear effects play a crucial role for characterizing phenomenological consequences of massive gravity. 
Motivated by this fact, the analysis of
   exact solutions of the equations of motion, obtained
 by imposing
 appropriate symmetries (spherical symmetry for static space-times, or homogeneity and isotropy for cosmological set-ups),
 make manifest, in idealized but representative situations, how the non-linear dynamics of the graviton
 degrees of freedom respond to the presence of a source, or, at very large scales,  to the  graviton mass
 itself.   This has been the argument  of
  this article,  in which we reviewed our works on these topics.   
   
   \smallskip
   
   Much interesting  work is left for the future:  
      our results can be extended in various directions that will improve our understanding of massive gravity and, in general, of
  consistent infrared modifications of General Relativity. From one side, it would be interesting to find new stationary configurations renouncing to spherical symmetry, to test analytically the effectiveness of Vainshtein mechanism when spherical symmetry is broken.
 As a concrete example, it would be interesting to find analogues  of the Kerr geometry in this scenario, in which frame dragging effects can be quantitatively analyzed.    
   Also, it would be interesting to determine cosmological configurations that break the isotropy or homogeneity of the
   cosmological solutions
     analyzed until now.  Indeed, working 
    in a suitable decoupling limit, 
   we have shown that the cosmological self-accelerating configurations studied so far 
   are characterized by instabilities in the vector sector.
    Given the recent results
     of \cite{DeFelice:2013awa}, 
      we speculate  that these instabilities can be possibly  avoided  by renouncing 
       to some of the symmetries that characterize the solutions (for example isotropy of the three spatial directions).
         It would be interesting to determine stable self-accelerating backgrounds 
       following this route, 
     and 
  study their consequences for what respect the dynamics of cosmological fluctuations.
  %
    % . It would be interesting to understand whether this feature is associated with the specific symmetries imposed on the system or it is a general feature.
  %and if so whether new stable and phenomenologically interesting configurations can be determined renouncing to these symmetries.
  We hope to be able to develp  all these questions in our future work.

\acknowledgments

We thank Fulvio Sbis\`a for collaboration on  part of the topics presented here.  
GT is supported by an STFC Advanced Fellowship ST/H005498/1. KK is supported by STFC grant ST/H002774/1, the ERC and the Leverhulme trust. GN is supported by the grants PROMEP/103.5/12/3680 and CONACYT/179208.

\begin{appendix}
\section{General exact solution}\label{AppA}

From the general Lagrangian (\ref{genlag}), and using the non-diagonal ansatz (\ref{genmetr}) together with Einstein equations (\ref{einsteineqns}), one can show that there are two branches of solutions for non-vanishing $\alpha_3$ and $\alpha_4$ as it was done in Section \ref{sphsols}. Here we only consider the branch with a non-diagonal metric, where analytic solutions can be found. Since the combination $\sqrt{1+3\alpha_3+9\alpha_3^2-12\alpha_4}$ is always present in the solution of this branch (see (\ref{solb0})), it is convenient to map the ($\alpha_3,\alpha_4$) parameters into ($\alpha_3,\alpha_5$), where $\alpha_5^2\equiv1+3\alpha_3+9\alpha_3^2-12\alpha_4$. In this new set of parameters, the combination  $d(r)\, G_{tt}+b(r)^2\,G_{tr}\,=\,0$, fixes $c(r)$ as a function of $r$ in the following way
\be\label{boapp}
c(r)= c_0 r \,=\,
\frac{(1+3 \alpha_3+\alpha_5)}{(2+3 \alpha_3+\alpha_5)} r.
\ee
The rest of Einstein equations give
\be\label{generalsol}
b(r)^2= \frac{\Delta_0}{c_0^2}(1-p), \qquad a(r)^2=\frac{\Delta_0}{c_0^2}(p+\gamma+1), \qquad d(r)=\sqrt{\Delta_0-a(r)^2b(r)^2},
\ee
where
\be\label{papp}
p=\frac{\mu}{r}+\frac{(1+3 \alpha_3+2 \alpha_5)}{3 (2+3 \alpha_3+\alpha_5)^2}m^2 r^2, \qquad
\gamma+1=\frac{(1+3 \alpha_3+\alpha_5)^4}{\Delta_0(2+3 \alpha_3+\alpha_5)^4}
\ee
Just like in the $\alpha_3=\alpha_4=0$ ($\alpha_5=1$) case, there are two integration constants, $\mu$ and $\Delta_0$, but in order to have a positive argument for the square root in $d(r)$, $\Delta_0$ has to run from $\Delta_0=0$ to $\Delta_0^{max}=c_0^2$. If we focus on the massless case $\mu=0$ only, the solution describes the static patch of the de Sitter or Anti-de Sitter spacetime.

\end{appendix}

\end{document}